# Bifurcation Analysis Framework of Spiking Neuron Models

Zhiwei Li, Shi-Li Zhang, Chenyu Wen

*Abstract*—**Neuromorphic computing targets energy-efficient event-driven information processing by placing artificial spiking-neurons at its core. Artificial neuron devices and circuits have multiple operating modes and produce region-dependent nonlinear dynamics that are not captured by system analysis methods for linear systems, such as transfer functions, Fourier/Laplace transform, Bode diagram, etc. Thus, new tools are needed to evaluate the nonlinear behavior of neurons and to guide the design and optimization of artificial neuron implementations. Here we present a generalized bifurcation analysis framework based on nonlinear dynamical systems theory. A CMOS axon–hillock neuron, memristor neuron, and the FitzHugh–Nagumo biological neuron model are selected for demonstration. We evaluate Hopf bifurcation conditions to define the rest and firing domains in parameter space of the system, and predict the near-onset firing rate. The results are further compared with numerical simulations. The framework standardizes the analysis for various neuron models and physical realizations. It yields practical design and optimization guidelines for artificial neurons for neuromorphic computing systems, including parameter combinations to make a neuron fire/rest and to control the corresponding firing properties, e.g., firing rate and amplitude.**

## I. INTRODUCTION

ARTIFICIAL intelligence (AI) is currently implemented predominantly in software on von Neumann architected computers, where physically separating memory and processing units induces large power and time costs in data transfer [1], [2], [3]. The resulting computation efficiency gap between current AI and human brain is stark: brain-comparable floating-point throughput realized in supercomputers consumes 20 MW of electrical power, 6 orders of magnitude higher than what human brains do [4], [5]. This mismatch has motivated neuromorphic computing, which aims to narrow the gap by replacing clocked arithmetic pipelines with networks of neuron-like processing elements and distributed memories [6], [7], [8], [9].

A defining feature of neuromorphic computing is that information is encoded in sequences of electrical pulses, i.e., spike trains, rather than by continuous voltage/current levels [10], [11]. In such systems, the timing and rate of spikes carry information, combining features of digital event occurrence with analog inter-spike intervals and enabling compute-in-memory and event-driven operations [12]. These properties promise substantial gains in energy efficiency and scalability. At the hardware level, implementing large-scale spiking neuron networks with well-controlled dynamics is key to realizing aforementioned advantages of neuromorphic computing, which in turn requires sophisticated design and optimization of artificial neurons at circuit and device levels [13].

In this context, neuromorphic hardware seeks to reproduce neuron-like spiking with controllable frequency, amplitude, and energy under device and circuit constraints [14], [15], [16], [17], [18]. In general, neuron networks are assembled by two elemental building blocks, neurons and synapses, implemented using CMOS circuits and, increasingly, emerging devices such as various memory devices or memristors [9], [19], [20], [21], [22]. At the neuron level, circuit designers must engineer spike onset, spike waveform, and firing dynamics while accommodating device nonidealities and fabrication variability [23]. Achieving these targets in a systematic way requires analysis tools that connect implementation choices to dynamic behaviors, rather than relying on ad hoc parameter sweeps and expert experience.

Regarding the neuron models, early efforts focused on compact phenomenological abstractions (e.g., integrate-and-fire, FitzHugh–Nagumo (FHN), and Morris–Lecar) and analog very-large-scale integration (VLSI) realizations, which demonstrated feasibility but offered limited guidance for engineering targets [24], [25]. With the emergence of neuromorphic devices, circuit–device co-design has become the dominant paradigm: nonlinearity, gain, and time constants are distributed across amplifiers, capacitors, transistors, and memory devices [26]. Traditional small-signal circuit analysis techniques (e.g., Fourier transform, Laplace transform, Bode plots, root locus, etc.) are inherently restricted to linear systems. Nonlinear circuits must be approximated by local linear models. Hence, global behaviors such as limit cycle generation and termination, as well as robustness to noise and variations in device characteristics, cannot be captured directly. This limitation motivates research to use tools from dynamical systems, which can treat nonlinear systems in a semi-quantitative topological manner. Neurons are oscillators

This work was partially financed by the faculty funds at Uppsala University. *(Corresponding author: Chenyu Wen).*

Zhiwei Li is with the Division of Solid-State Electronics, Department of Electrical Engineering, Uppsala University, SE-751 21 Uppsala, Sweden.

Shi-Li Zhang was with Division of Solid-State Electronics, Department of Electrical Engineering, Uppsala University, SE-751 21 Uppsala, Sweden.

Chenyu Wen is with the Division of Solid-State Electronics, Department of Electrical Engineering, Uppsala University, SE-751 21 Uppsala, Sweden (e-mail: chenyu.wen@angstrom.uu.se).

Color versions of one or more of the figures in this article are available online at http://ieeexplore.ieee.org

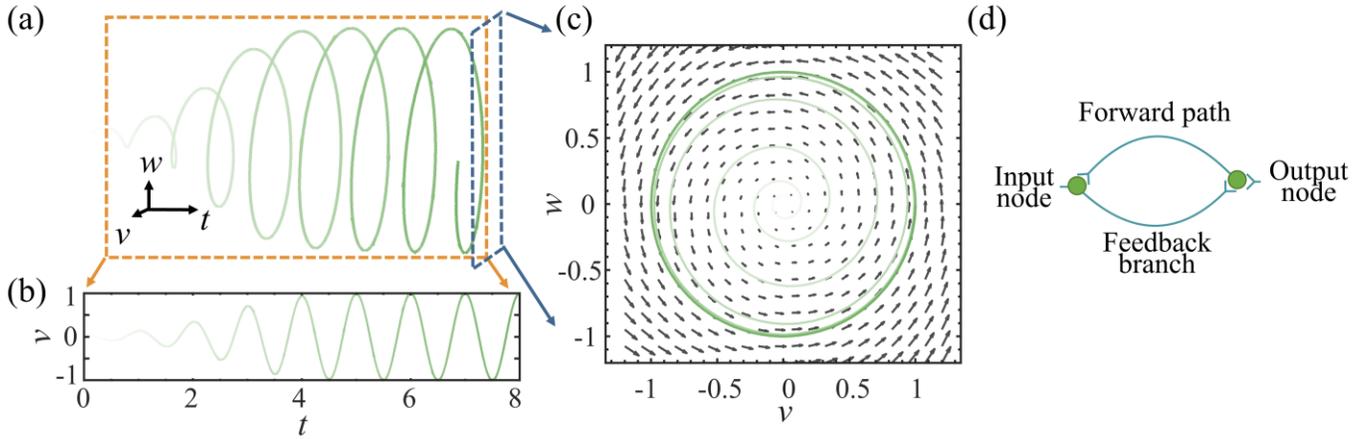

**Fig. 1.** Dynamical systems view of a firing neuron. (a) Three-dimensional state–time view showing the evolution of the trajectory over time. (b) Evolution of representative state variable, $v$, in time domain. (c) Phase portrait depicting the state space projection without explicit time. (d) System block diagram of a spiking neuron.

whose states (rest or firing) are determined by fixed points of the system as well as bifurcations of their governing equations.

As illustrated in Fig. 1, a simple firing neuron can be modeled by a two-dimensional (2D) system whose state $(v, w)$ can be represented as points on state plane. The state evolution with time, $t$, is illustrated in Fig. 1(a) as an example, and the resulting spike train in the time domain as the variable $v(t)$ is shown in Fig. 1(b). From another aspect, the state evolution of the system can be projected as trajectories in the state space, named phase portraits, as shown in Fig. 1(c). The vector field in the state space indicates the change direction and tendency of the state, *i.e.*, the "speed" of the state, at corresponding point along time. This change of viewpoint, from waveforms in time domain to trajectories in state space, underpins the dynamical systems approach: starting from ordinary differential equations (ODEs) of the systems, one analyzes state evolution through phase portraits, nullclines, and bifurcation structure to identify fixed points and their stability, firing domain, and trajectories. A large body of research has applied this approach to analyze neuronal dynamics, as well as the model-parameter dependence of spiking properties for different artificial neurons [27], [28], [29], [30]. However, the conclusions thus far remain case specific because they are typically performed on individual models/realizations. Moreover, these analyses are typically performed by numerically solving the neuron models. Although the numerical solutions can capture key dynamic features of neurons, they obscure the overview of the nonlinear structure of the neuron models/circuits. Important causes of the nonlinearity in neuron circuits are usually due to amplifier saturation, transistor operating mode transitions, and memristor state shifting. Most existing dynamical analyses do not explicitly address the associated transition of operating regions of the devices/systems in state space, thereby obscuring the mechanisms by which oscillations emerge or disappear near these transition boundaries [31], [32], [33]. Furthermore, model parameters are rarely normalized or benchmarked across platforms, hindering comparisons between different neuron models, and making it difficult to transpose design insights [34]. Collectively, these limitations impede a systematic exploration of device and circuit parameter selection and optimization, obstruct systematic tuning of spiking properties, and complicate transferable hardware design [35], [36], [37], [38].

Here, we develop a generalized analysis framework for neuron behavior based on dynamical systems analysis that directly addresses the limitations outlined above. Methodologically, the framework is based on dynamical systems theory, including Hopf bifurcation, and is formulated so as to not rely on any particular circuit structure or artificial neuron realization. We first describe various neuron realizations based on a common 2D system with forward path and feedback branch. Given the 2D system and its corresponding operating mode of devices, the framework (i) divides operating region in state space, (ii) computes fixed points and their stability strictly within each operating region, (iii) identifies Hopf onset and near-onset firing rate from the Jacobian, while (iv) compares the firing rate with numerical simulation results. This workflow replaces the current unstructured parameter sweep analysis with a system-level view that is independent of the underlying realization and directly links device and circuit parameters to dynamic behaviors of neurons.

A central goal of the framework is to answer questions in neuron design in a manner useful for hardware engineers. By normalizing circuit/device parameters, the method exposes how the interplay of the dynamics of slow- and fast-changing nodes in each operating region in the state space governs the appearance and robustness of spiking. It further reveals how a certain device/component, such as those introducing a feedback capacitor, relocating a delay element from a forward path to a feedback branch, or changing leak conductance, modifies, influences, or reshapes operating regions, spike onset, and spike waveform and frequency.

The remainder of the paper is arranged as follows: Section 2 presents the analysis workflow, model formulation, normalization, and bifurcation conditions. Section 3 illustrates the workflow for three representative applications, i.e., an axon–hillock neuron, a memristor neuron, and the FHN model. Section 4 summarizes the work with potential extensions of the methodology and workflow to other applications.

## II. METHODS

### 2.1 Bifurcation analysis workflow

Regarding the behavior of a neuron, as schematically shown in Fig. 1(d), it can be modeled as a two-node system that receives stimulus signal as input and produces spikes as output. The input, typically an electrical current, charges a "membrane" capacitance as the input node. It integrates the input over time reflected as the voltage on the capacitance, denote by $v$. When $v$ reaches a threshold, the neuron emits a spike from the output node with an electrical potential $w$. Then, a feedback branch discharges the capacitor and resets the state of input node so that the integration restarts. This integrate-and-fire behavior can be mathematically described by ODEs about $v$ and $w$ as,

$$\frac{dv}{dt} = f(v, w, \theta)$$
$$\frac{dw}{dt} = g(v, w, \theta)$$
(1)

where, $\theta$ collects controllable parameters of the system (e.g., bias currents, conductance, capacitances, thresholds, etc.). Two state variables are sufficient to capture spike generation and recovery in the conceivable architectures. Here, $v$ plays the role of a membrane-like variable at the input node, while $w$ of the output node controls the feedback. In hardware realizations, (1) is often piecewise-defined because amplifiers, transistors, and memory devices switch between different operating modes, which are described by different mathematical functions.

On the $(v, w)$ plane, the $v$ and $w$ nullclines of (1) are defined by,

$$\mathcal{N}_v: f(v, w, \theta) = 0$$
$$\mathcal{N}_w: g(v, w, \theta) = 0$$
(2)

They are curves on which $v$ or $w$ is instantaneously stationary with respect to time, i.e., $dv/dt$ or $dw/dt$ is zero on the nullclines. Their intersections, $x^* = (v^*, w^*)$, give fixed points where both derivatives vanish, and the phase portrait indicates how the system state approaches or departs from the fixed points from the initial states, shown as the trajectories in the state space.

To characterize how the system behaves in the neighborhood of a fixed point $x^* = (v^*, w^*)$, the system is linearized around $x^*$. Assume that $f$ and $g$ are smooth and the eigenvalues of the Jacobian at $x^*$ do not have zero real part, the linearization (Hartman–Grobman) theorem [39] ensures that, within a certain neighborhood of $x^*$, the linearized system is qualitatively equivalent to the original system. In other words, the original system can be approximated by a linearized system near the fixed points. In practice, this means that the eigenvalues of the Jacobian describe how trajectories in a finite neighborhood of $x^*$ approach (stable) or depart from

(unstable) the fixed point and whether they do so monotonically or in an oscillatory fashion. The nature of the trajectories can be inferred from the Jacobian matrix evaluated at $x^*$,

$$J = \begin{bmatrix} \dfrac{\partial f}{\partial v} & \dfrac{\partial f}{\partial w} \\ \dfrac{\partial g}{\partial v} & \dfrac{\partial g}{\partial w} \end{bmatrix}$$
(3)

Beyond the local classification of individual fixed points, it is important to understand how the behavior of the system qualitatively (topologically) changes as a parameter in $\theta$ is varied. Such qualitative changes are called bifurcations: a bifurcation occurs when a change in a parameter alters the number or stability type of fixed points or periodic orbits. In the neuron models considered here, the primary local bifurcation of interest is the Hopf bifurcation (also known as Poincaré–Andronov–Hopf bifurcation) [40]. At a Hopf bifurcation point, a fixed point changes stability as a complex-conjugate pair of eigenvalues of the Jacobian crosses the imaginary axis, and a periodic trajectory, named limit cycle, appears or vanishes. For a neuron, varying system parameters such as bias current, leak conductance, and feedback properties may cause Hopf bifurcations, which indicates a transition between a quiescent operating point and sustained repetitive firing.

For the conditions under which the Hopf bifurcation happens (see Section 2.5 for specific conditions), the parameter combinations that make a neuron firing or rest can be derived. Furthermore, the firing frequency near the onset can be estimated. These results can explicitly answer relevant questions in design and optimization of a neuron. For example, for a given circuit topology, which parameter combinations yield only rest (i.e., the rest domain, RD, in the parameter space), which induce stable spiking (i.e., the firing domain, FD, in the parameter space), how the near-onset firing rate scales with key circuit/device parameters. By tracking how FD changes when, for example, a feedback capacitor is added or a load conductance is changed, it provides system level insights into the way that topology and parameters shape neuronal behaviors.

### 2.2 ODEs of a neuron model

Here, the most commonly used CMOS axon–hillock neuron [Fig. 2(a)] is employed as an example to demonstrate the analysis in the proposed framework above. It is worth noting that any other neuron devices, circuits, or systems can be analyzed by following a similar procedure. In the *Results and discussion* section, analysis results will be shown for a memristor neuron model and a biological neuron model using this method. By applying the Kirchhoff current law for the input node and the Kirchhoff voltage law for the feedback loop, the 2D model is written as,

$$(C_{\text{mem}} + C_{\text{f}})\frac{dv}{dt} = I_{\text{in}} - g_{\text{L}}v$$
$$+ \frac{C_{\text{f}}}{\tau_A}[w_\infty(v, \theta_A) - w] - I_{fb}(v, w, \theta_T) \quad (4)$$

$$\tau_A \frac{dw}{dt} = w_\infty(v, \theta_A) - w$$

As previously defined, $v$ is the input node voltage and $w$ the output node voltage. The total capacitance at $v$ is $C_{\text{mem}} + C_{\text{f}}$. $g_{\text{L}}$ is the leak conductance given by $g_{\text{L}} = \frac{1}{R_{\text{L}}}$, $I_{\text{in}}$ the applied current drive, $\tau_A$ the delay time of the amplifier, and $\theta_A$ and $\theta_T$ parameter sets of the amplifier and transistor, respectively. It is worth noting that the delay section can also be implemented in the feedback branch, which can be equivalent to the amplifier-delay configuration shown here and similarly analyzed. The derivation of (4) is detailed in Note 1 of Supporting Information (SI).

The output characteristics of a MOS field-effect transistor is described by [41],

$$I_{\text{fb}}(v, w, \theta_{\text{M}}) =$$
$$\begin{cases} 0, & w < V_{\text{gth}} \\ \frac{k}{2}\left(w - V_{\text{gth}}\right)^2, & w > V_{\text{gth}}, v > w - V_{\text{gth}} \\ k\left(w - V_{\text{gth}}\right)v - \frac{k}{2}v^2, & w > V_{\text{gth}}, v < w - V_{\text{gth}} \end{cases} \quad (5a)$$

where, $k$ is an effective transconductance parameter and $V_{\text{gth}}$ the threshold voltage. The corresponding characteristics is shown in Fig. S1 of the SI.

The transfer characteristic of the amplifier is approximated by

$$w_\infty(v, \theta_A) = \begin{cases} V_1, & v < V_{\text{th1}} \\ \frac{v - V_{\text{th1}}}{(V_{\text{th2}} - V_{\text{th1}})}V_{\text{dd}}, & V_{\text{th1}} < v < V_{\text{th2}} \\ V_2, & v > V_{\text{th2}} \end{cases} \quad (5b)$$

where, $V_1$ and $V_2$ are, respectively, the lower- and upper-bound output levels. When $v$ is smaller than $V_{\text{th1}}$, the output is $V_1$, while when $v$ is larger than $V_{\text{th2}}$, the output is $V_2$. In between $V_{\text{th1}}$ and $V_{\text{th2}}$, the output linearly increases from $V_1$ to $V_2$. A typical choice is $(V_1, V_2) = (0, V_{\text{dd}})$, as illustrated in Fig. S2 of the SI. It is worth noting that to attain a closed-form analytical solution for the fixed point, a linear segmented function is adopted for the amplifier characteristics, instead of a S-shaped curve (e.g., a sigmoid function). In this way, the fixed point can be explicitly expressed as the solutions of a polynomial equation group about $w$ and $v$ lower than the order of three. However, this approximation may cause errors and the polynomial order should not be too low, which will be discussed later.

The lines $v = V_{\text{th1}}$, $v = V_{\text{th2}}$, $w = V_{\text{gth}}$ and $v = w - V_{\text{gth}}$ define the boundaries on the $(v, w)$ plane They divide the state space into 9 different regions by three operating modes of the amplifier, $w \in (-\infty, V_1], (V_1, V_2), [V_2, \infty)$, in combination with three operating states of the transistor, Off, Saturation, and Linear. The corresponding expressions of $w_\infty(v, \theta_A)$ and $I_{\text{fb}}(v, w, \theta_{\text{M}})$ for each region are applied.

The amplifier is modeled as a first-order delay by the second formula of (4) that drives $w$ toward its equilibrium value $w_\infty$. In terms of the $v$-node, its contribution is $\frac{C_{\text{f}}}{\tau_A}(w_\infty - w)$ by the feedback capacitor $C_{\text{f}}$. Another feedback branch is implemented by a grounded-source n-type MOSFET whose current is given by (5a).

Within each of the 9 regions, the vector fields, $f(v, w, \theta)$ and $g(v, w, \theta)$, are smooth. At the boundaries between these regions, the vector field is continuous. Therefore, the trajectories of the system state are globally smooth.

### 2.3 Normalization and non-dimensionalization

To non-dimensionalize parameters, the variables are, respectively, normalized to voltage $V_{\text{dd}}$, time $\tau_A$, and unit charge $q$ ($1.6 \times 10^{-19}$ C). Hence, $\hat{v} = \frac{v}{V_{\text{dd}}}$, $\widehat{w} = \frac{w}{V_{\text{dd}}}$, $\hat{t} = \frac{t}{\tau_A}$, $\widehat{C}_{\text{f}} = \frac{C_{\text{f}}V_{\text{dd}}}{q}$, $\widehat{I_{\text{in}}} = \frac{I_{\text{in}}\tau_A}{q}$, $\widehat{g_{\text{L}}} = \frac{g_{\text{L}}\tau_A V_{\text{dd}}}{q}$, $\widehat{V_{\text{th1}}} = \frac{V_{\text{th1}}}{V_{\text{dd}}}$, $\widehat{V_{\text{th2}}} = \frac{V_{\text{th2}}}{V_{\text{dd}}}$, $\widehat{V_{\text{gth}}} = \frac{V_{\text{gth}}}{V_{\text{dd}}}$, and $\hat{k} = \frac{k\tau_A V_{\text{dd}}^2}{q}$. In the model, the dimensionless parameters are $\hat{\theta} = \{\widehat{I_{\text{in}}}, \widehat{g_{\text{L}}}, \hat{k}, \widehat{C}_{\text{f}}, \widehat{V_{\text{th1}}}, \widehat{V_{\text{th2}}}, \widehat{V_{\text{gth}}}\}$. The following derivation uses the operating region where the amplifier is located at $\widehat{V_{\text{th1}}} < \hat{v} < \widehat{V_{\text{th2}}}$ whereas the transistor is set in the Linear mode $\hat{v} < \widehat{w} - \widehat{V_{\text{gth}}}$ as a representative case. The other regions can be treated in a same fashion and can be found in Note 1 of the SI. From (4), the normalized system ODEs in this region are,

$$\frac{d\hat{v}}{d\hat{t}} = \frac{1}{2\widehat{C}_{\text{f}}}[\widehat{I_{\text{in}}} - \widehat{g_{\text{L}}}\hat{v} - \left(\hat{k}(\widehat{w} - \widehat{V_{gth}})\hat{v} - \frac{\hat{k}}{2}\hat{v}^2\right)$$
$$+ \widehat{C}_f\left(\frac{\hat{v} - \widehat{V_{th1}}}{(\widehat{V_{th2}} - \widehat{V_{th1}})} - \widehat{w}\right)] \quad (6a)$$

$$\frac{d\widehat{w}}{d\hat{t}} = \frac{\hat{v} - \widehat{V_{th1}}}{(\widehat{V_{th2}} - \widehat{V_{th1}})} - \widehat{w} \quad (6b)$$

This non-dimensionalization allows results from different parameter choices, operating regions, and physical realizations to be compared and interpreted in a consistent way.

### 2.4 Fixed points and linearization

Likewise, the nullclines of the system from (2) are,

$$\widehat{\mathcal{N}_v}: \widehat{I_{\text{in}}} - \widehat{g_{\text{L}}} \cdot \hat{v} + \widehat{C}_{\text{f}} \cdot (\widehat{w_\Delta} - \widehat{w}) - \widehat{I_{\text{fb}}}(\hat{v}, \widehat{w}, \widehat{\theta_{\text{M}}}) = 0$$
$$\widehat{\mathcal{N}_w}: \widehat{w_\infty}(\hat{v}, \widehat{\theta_A}) - \widehat{w} = 0 \quad (7)$$

which give immediate geometric insights: increasing $\widehat{I_{\text{in}}}$ shifts $\widehat{\mathcal{N}_v}$ upward while increasing $\widehat{g_{\text{L}}}$ steepens the negative slope of $\widehat{\mathcal{N}_v}$ (cf. Fig. 2(b)).

Here, the fixed points are the intersections of curves $\widehat{\mathcal{N}_v}$ and $\widehat{\mathcal{N}_w}$ given by,

$$\widehat{I_{\text{in}}} - \widehat{g_{\text{L}}}\hat{v}^* - \left(\hat{k}(\widehat{w}^* - \widehat{V_{\text{gth}}})\hat{v}^* - \frac{\hat{k}}{2}\hat{v}^{*2}\right) = 0$$
$$\frac{\hat{v}^* - \widehat{V_{\text{th1}}}}{(\widehat{V_{\text{th2}}} - \widehat{V_{\text{th1}}})} - \widehat{w}^* = 0 \quad (8)$$

Solving (8) yields

$$\hat{v}^* = \begin{cases} \hat{k}\left(\dfrac{V_{th1}}{\Delta \hat{V}} + \widehat{V_{gth}}\right) - \widehat{g_L} \\ \pm \sqrt{\left[\widehat{g_L} - \hat{k}\left(\dfrac{V_{th1}}{\Delta \hat{V}} + \widehat{V_{gth}}\right)\right]^2 + 4\hat{k}\left(\dfrac{1}{\Delta \hat{V}} - \dfrac{1}{2}\right)\widehat{I_{in}}} \end{cases} / 2\left(\dfrac{\hat{k}}{\Delta \hat{V}} - \dfrac{\hat{k}}{2}\right)$$

$$\hat{w}^* = \frac{\hat{v}^* - V_{th1}}{\Delta \hat{V}}$$

where, $\Delta \hat{V} = \widehat{V_{th2}} - \widehat{V_{th1}}$. An additional validation should be made on the position of these calculated fixed points, which should lie in the corresponding operating regions.

Local stability follows from the Jacobian of (6) evaluated at $(\hat{v}^*, \hat{w}^*)$:

$$J(\hat{v}^*, \widetilde{w}^*) = \begin{bmatrix} \dfrac{1}{2\hat{C}_f}\left(-\widehat{g_L} + \dfrac{\hat{C}_f}{(\widehat{V_{th2}} - \widehat{V_{th1}})}\right) & -\dfrac{1}{2} \\ \dfrac{1}{(\widehat{V_{th2}} - \widehat{V_{th1}})} & -1 \end{bmatrix}_{(\hat{v}^*, \hat{w}^*)} \quad (9)$$

The fixed point is stable if $Re\,(\lambda_{1,2}) < 0$ and unstable if at least one eigenvalue satisfies $Re(\lambda_{1,2}) > 0$, where $\lambda_{1,2}$ are the eigenvalues of $J(\hat{v}^*, \hat{w}^*)$.

### 2.5 Local bifurcations and criteria

The trace and determinant of the Jacobian are, respectively,

$$Tr = \frac{1}{2\hat{C}_f}\left(-\widehat{g_L} + \frac{\hat{C}_f}{(\widehat{V_{th2}} - \widehat{V_{th1}})}\right) - 1$$

$$\Delta = \frac{-1}{2\hat{C}_f}\left(-\widehat{g_L} + \frac{\hat{C}_f}{(\widehat{V_{th2}} - \widehat{V_{th1}})}\right) + \frac{1}{2(\widehat{V_{th2}} - \widehat{V_{th1}})} \quad (10)$$

In 2D linearized systems, $Tr$ and $\Delta$ completely govern the local (codimension-one) bifurcations of the fixed points [39]. The common cases in a 2D system are:

• Saddle–node: $\Delta = 0$ and $Tr \neq 0$ (generic nondegeneracy).

• Hopf: $Tr = 0$ and $\Delta > 0$, with $dTr/d\theta_i \neq 0$ for certain parameter $\theta_i \in \{\widehat{I_{in}}, \widehat{g_L}, \hat{k}, \hat{C}_f, \widehat{V_{th1}}, \widehat{V_{th2}}, \widehat{V_{gth}}\}$.

For a supercritical Hopf bifurcation that indicates the oscillation onset [39] further requires the first Lyapunov coefficient [39]

$$Lv = \frac{1}{16}(f_{vvv} + f_{vww} + g_{vvw} + g_{www})$$
$$+ \frac{1}{16\omega}(f_{vw}(f_{vv} + f_{ww}) - g_{vw}(g_{vv} + g_{ww})$$
$$- f_{vv}g_{vv} + f_{ww}g_{ww}),$$

to be negative, where, $\omega = Im\,(\lambda_{1,2})$ and $f_{vw} = \frac{\partial^2 f}{\partial v \partial w}$ (with similar definitions for the other partial derivations). In this way, a limit cycle is generated and the oscillation period near onset satisfies,

$$T(\theta_0) \approx \frac{2\pi}{\omega_0}, \quad (11)$$

where, $\pm i\omega_0$ are the eigenvalues of $J(\hat{v}^*, \hat{w}^*)$ in (9) at the bifurcation point, $\theta_i = \theta_{i,0}$.

In summary, a neuron firing through the Hopf bifurcation must satisfy the following conditions on the system parameters, often named the Hopf conditions, through,

1. Calculating all the fixed points of the system in each segment on the $(v, w)$ plane. There should be at least one fixed point.

2. Calculating the Jacobian matrix at these fixed points, and at least one fixed point should possess a complex conjugated eigenvalue pair $a \pm i\hat{w}$. Then, $Tr = 2a$ and $\Delta = a^2 + \omega^2$.

3. For certain parameter $\theta_i$ that $a$ depends on, varying it to cause $a$ change sign, $i.e.$, from positive to negative and vice versa. Finding the critical value $\theta_i = \theta_{i,0}$ that makes $a = 0$, $i.e.$, $Tr = 0$, which can be a possible critical point of the Hopf bifurcation.

4. Computing $\Delta$ at $\theta_i = \theta_{i,0}$ and making sure that it assumes a positive value.

5. Computing $dTr/d\theta_i$ at $\theta_i = \theta_{i,0}$ and making sure that it is non-zero.

6. Computing $Lv$ at $\theta_i = \theta_{i,0}$ and making sure that it is a negative value.

These conditions give a group of inequalities of the parameters that make a neuron fire (see Note 2 of the SI for more details). In addition, each operating region in the parameter space corresponds to a unique device operating mode and is specified by a set of inequality constraints (see Note 1 of the SI for details). The FD in the parameter space is further confined by the boundaries of each operating region. Thus, the total FD in the parameter space of the system is the unification of those from each operating region.

### 2.6 Periodic orbits: simulation and validation

To verify the theoretical predictions, numerical simulations of the system were implemented in MATLAB (the simulation scripts are provided in SI). In the simulations, a trajectory was classified as periodic when two successive returns differed by less than a prescribed tolerance, meaning that the state mismatch was smaller than $\varepsilon$ in norm and the relative change in the return time was also smaller than $\varepsilon$. The period was measured as the time difference between two consecutive section crossings, $T$. Thus, the firing rate is $f = 1/T$. The amplitude is $A = \max_{t\in[0,T]} \hat{v}(t) - \min_{t\in[0,T]} \hat{v}(t)$.

Though simulations, two points are verified: 1) If the system transition from rest to firing as corresponding parameters vary is predicted by the theory and 2) In the firing state, if the firing rate matches that predicted by the theory.

## III. RESULTS AND DISCUSSION

### 3.1 CMOS axon–hillock neuron

By inserting typical values of the circuit parameters in the model, the nullclines and fixed points of the axon–hillock neuron are calculated in Fig. 2 (b). The typical values of the parameters used here are: $C_f = 1$ nF, $k = 1 \times 10^{-5}$ A/V$^2$, $V_{th1} = 0.5$ V, $V_{th2} = 1.0$ V, $V_{dd} = 3.0$ V, $V_{gth} = 1.5$ V, $g_L = 1 \times 10^{-9}$ S, $T_A = 0.4$ ms, $C_{mem} = 1$ nF, and $I_{in} = 5.15\ \mu$A. Three different operating modes of the amplifier and transistor divide the parameter plane into 9 different operating regions, with the boundaries shown as the dash lines in Fig. 2(b). Furthermore, the $w$- and $v$- nullclines can be calculated (solid lines in Fig. 2(b)) and the interception defines one fixed point located at $(0.92, 2.52)$. By calculating the Jacobian matrix at this point using (9), the eigenvalues of the matrix are found to be $2250.5 \pm 8073i$. Since the real part is larger than zero, this fixed point is unstable, which suggests a possible oscillation, *i.e.*, firing. With the initial point at $(0.1, 0.1)$, the trajectory of system evolution governed by the vector field of the system indicates a firing state as shown in Fig. 2(c).

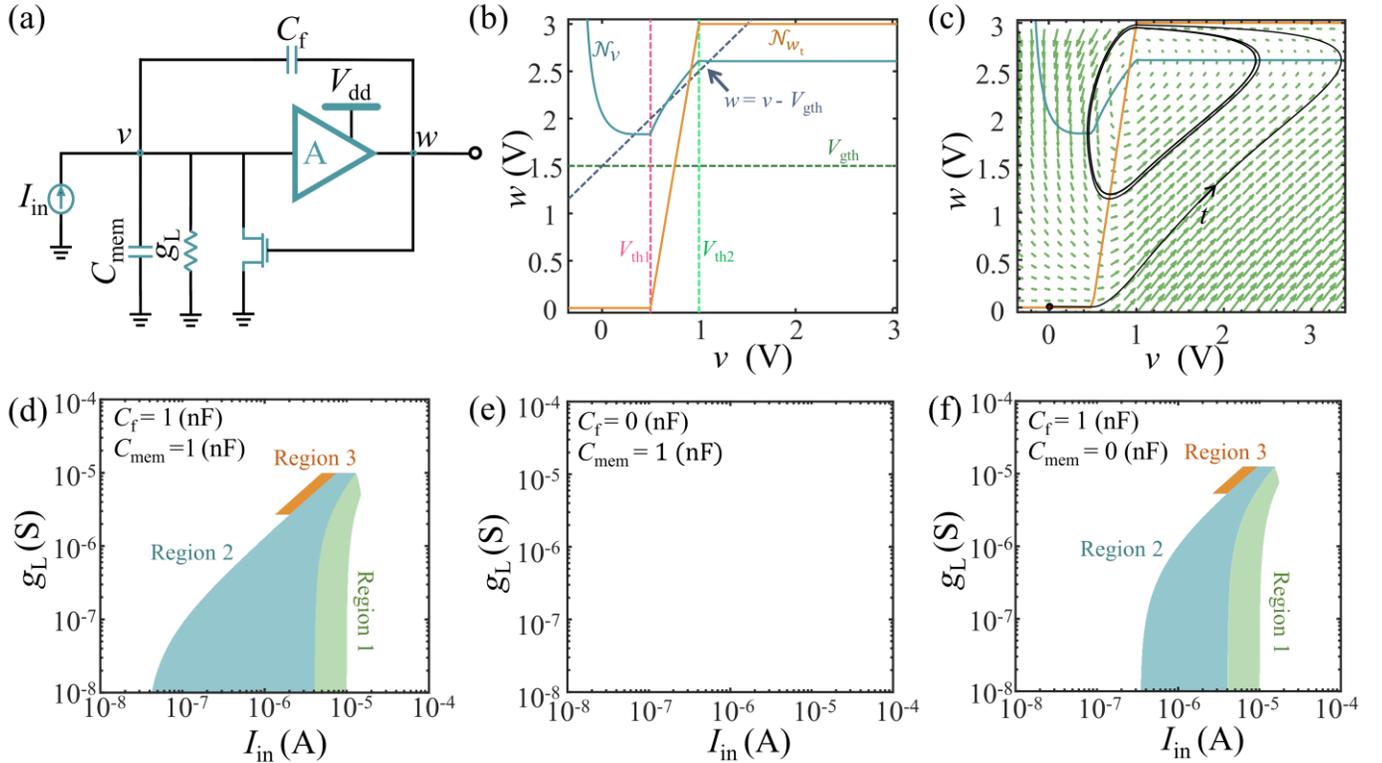

**Fig. 2.** Dynamic behaviors of the CMOS axon–hillock neuron. (a) Neuron circuit schematic realized by using CMOS devices. (b) The $v$- and $w$-nullclines (solid curves) and operating region boundaries (dashed lines) in $(v, w)$ state plane. (c) Phase portrait trajectory from the marked starting point; arrows denote the vector field (vector field arrows) and the black curve is the trajectory. (d–f) The firing domain in $(I_{in}, g_L)$ plane with different values of $C_{mem}$ and $C_f$. The firing domains from different operating regions are shown in different colors. See Note S1 in the SI for the conditions of each operating regions. All other parameters are identical to those used in the main text.

During the design and optimization of the circuit, it is critical to select a group of parameters that make the neuron fire as expected. For example, for a given neuron circuit, which combinations of input bias $I_{in}$ and leak conductance $g_L$ would support firing rather than leaving the neuron rest? Following the analysis framework, $I_{in}$ and $g_L$ can be set free while leaving the rest parameters fixed. Then, the inequality group from the Hopf conditions (see above under Section 2.5) and the operating region boundaries jointly define the FD in the parameter space $(I_{in}, g_L)$, as shown in Fig. 2(d). All the combinations of $I_{in}$ and $g_L$ in the FD will guarantee a firing neuron.

Furthermore, an important question for neuron design is whether a feedback capacitor, $C_f$, is necessary for neuron firing and, if so, how its value and the membrane capacitance, $C_{mem}$, would shape the FD. It can be observed that if the feedback capacitor is removed, $C_f = 0$, none of the parameter combinations in the study range can induce a firing neuron. In this case, the dynamics is in its rest state and the trajectories converge to an interior fixed point (see Fig. S3 of the SI). However, if the membrane capacitor is removed ($C_{mem} = 0$), spiking onsets for a subset of parameter combinations, but the FD in the parameter space shrinks compared to that with a membrane capacitor (*cf*., Fig. 2(e)).

Introducing a finite $C_f$ restores a two-timescale structure: the input node integrates input signals. For a small $C_{mem}$, this system has a narrow FD on the $(I_{in}, g_L)$ plane within which the system exhibits a stable limit cycle (Fig. 2(f)). Outside this domain, the dynamics returns to a fixed point. With increasing $C_{mem}$ (e.g., to 1 nF), the FD broadens and shifts toward lower input bias (Fig. 2(d)). In a dynamic perspective, a larger capacitor indicates a larger time constant of delay that would lengthen the portions of the trajectory in the phase portrait associated with $v$, allowing the membrane capacitor to be charged and discharged in a way that would sustain firing over a wider range of $(I_{in}, g_L)$. Fig. 2(e–f), together with Fig. S4 in the SI, illustrate these changes in the FD as $(k, C_{mem})$ are varied. In addition, the FD can be calculated and presented on other parameter planes, *e.g.*, $(k, V_{gth})$, $(I_{in}, C_f)$, or $(C_f, C_{mem})$, which can be found in Fig. S5 and S6 of the SI.

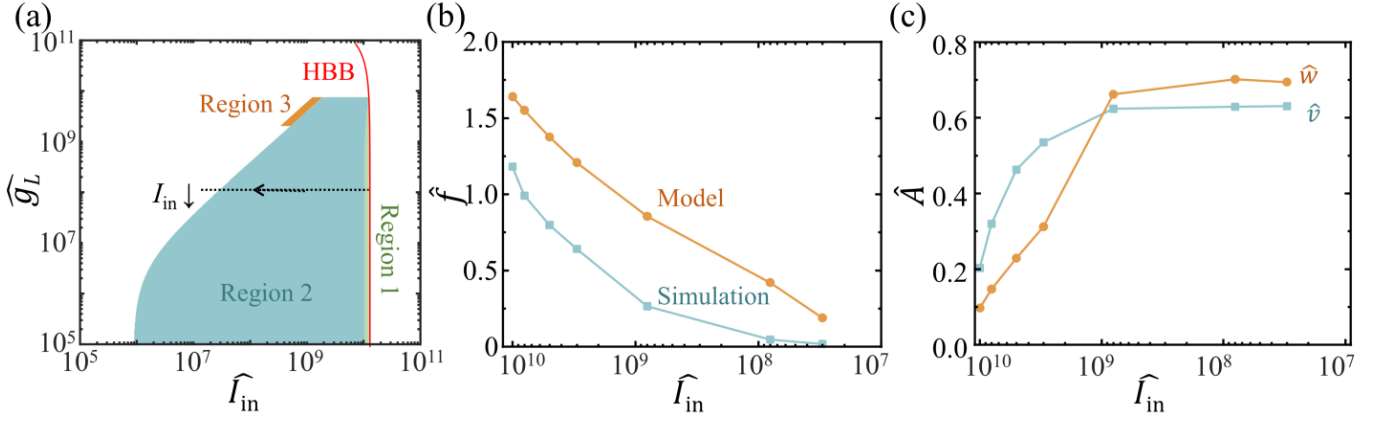

**Fig. 3**. Firing properties of the CMOS axon–hillock neuron. (a) The dash line marks the selected parameters used to assess the effect of $\widehat{I_{in}}$ on firing properties, that $\widehat{I_{in}}$ decreases from HBB to the interior while $\widehat{g_L}$ is kept constant ($\widehat{g_L} = 10^8$). The red curve marks HBB. All other parameters are identical to those used in the main text. (b) Firing rate changes along $\widehat{I_{in}}$ from model and simulation. (c). Amplitudes of $\hat{v}$ and $\hat{w}$ changes along $\widehat{I_{in}}$.

The spike amplitude $\hat{A}$ and firing rate $\hat{f}$ at the selected points in the FD on the $(\widehat{I_{in}}, \widehat{g_L})$ plan with changing $\widehat{I_{in}}$ are evaluated in Fig. 3. Points of interest are selected close to the Hopf bifurcation boundary (HBB) defined by $\widehat{Tr} = 0$, as indicated in Fig. 3(a). Moving away from this boundary by decreasing $\widehat{I_{in}}$ increases $\hat{A}$ and decreases the firing rate. The trajectory explores regions in the state space where the vector field is weak. Therefore, the system spends longer time on the slow recovery process and the waveform becomes more asymmetric with a steeper rise but a more gradual decay. Consequently, the mean firing rate decreases.

The firing rate obtained from the numerical simulation is compared with the near-onset firing rate predicted by the theoretical model in Fig. 3(b). The two agree closely near HBB, confirming that the local linearization approximates the simulated system efficiently and captures the firing rate at the onset. As $\widehat{I_{in}}$ decreases, the trajectory steps into the regions with a weaker vector field and the simulated firing rate increases more rapidly than the prediction from the linearized system based on the theoretical model that provides a lower bound for the firing rate.

The deviation itself is informative about how nonlinearity and boundary affect the slowdown of firing. Finally, how the amplitude of the input and output nodes, $\hat{v}$ and $\hat{w}$, changes with decreasing $\widehat{I_{in}}$ is displayed in Fig. 3(c). Close to HBB, the normalized voltages $\hat{v}$ and $\hat{w}$ vary almost proportionally to each other, reflecting the fact that the dynamics are controlled by the local linearization at the fixed point. As $\widehat{I_{in}}$ is reduced further, $\hat{w}$ grows more rapidly than $\hat{v}$ because the input node, $v$, saturates earlier in the firing cycle, which steepens the rising edge of the spike and increases the spike amplitude.

### 3.2 Memristor neuron

To demonstrate the capability of the generalization, the developed framework is applied to a memristor neuron in which the delayed feedback is implemented by a volatile, diffusive memristor, as shown in Fig. 4(a). In this realization, the voltage on the output node is again the system variable $w$, while the conductance of the memristor is reflected by the internal state $x \in [0,1]$ of the memristor, which acts as a slow-changing variable. The ODEs of the system are,

$$\frac{dx}{dt} = -u(x) - qw$$
$$\tau \frac{dw}{dt} = V_{ext} - \left[1 + \frac{R_{ext}}{R(x)}\right]w \quad (13)$$

where, $q > 0$ determines the strength of the conductance changes induced by the voltage, $\tau > 0$ is the characteristic timescale of the forward path defined by $\tau = R(x)C_f$, and $V_{ext}$ and $R_{ext}$ represent the input voltage and load resistance. $u(x)$ is the leakage of memristor, which can be described by a segmented function with thresholds $x_{th1}$ and $x_{th2}$,

$$u(x) =$$
$$\begin{cases} xk_1 + x_{u0}, x < x_{th1} \\ (x - x_{th1})k_2 + x_{th1}k_1 + x_{u0}, x_{th1} < x < x_{th2} \\ (x - x_{th2})k_3 + (x_{th2} - x_{th1})k_2 + x_{th1}k_1 + x_{u0}, x_{th2} < x \end{cases} \quad (14)$$

The resistance of the memristor, $R(x)$, is defined as a segmented function with thresholds $x_{rth1}$ and $x_{rth2}$,

$$R(x) =$$
$$\begin{cases} xk_{r1} + 1, x < x_{th1} \\ (x - x_{rth1})k_{r2} + x_{rth1}k_{r1} + 1, x_{th1} < x < x_{th2} \\ (x - x_{rth2})k_{r3} + (x_{rth2} - x_{rth1})k_{r2} + x_{rth1}k_{r1} + x_{u0}, x_{th2} < x \end{cases} \quad (15)$$

The plots of $u(x)$ and $R(x)$ can be found in Fig. S7 of the SI. The boundaries of the device mode $\{x_{th1}, x_{th2}, x_{rth1}, x_{rth2}\}$ define 5 different operating regions in the state space of the system (see Note 3 of the SI for details). In total, this model possesses 15 parameters, $\theta = \{x_{th1}, x_{th2}, x_{u0}, k_1, k_2, k_3, q, \tau, V_{ext}, R_{ext}, x_{rth1}, x_{rth2}, k_{r1}, k_{r2}, k_{r3}\}$. Unless otherwise stated, the following baseline parameters are used for the memristor neuron model, $x_{th1} = 0.1$, $x_{th2} = 0.95$, $x_{u0} = -8 \text{ s}^{-1}$, $k_1 = 80 \text{ s}^{-1}$, $k_2 = -0.2353 \text{ s}^{-1}$, $k_3 = 200 \text{ s}^{-1}$, $q = 1.265 \text{ V}^{-1}\text{s}^{-1}$, $\tau = 30 \text{ s}$, $V_{ext} = 0.5 \text{ V}$, $R_{ext} = 600 \Omega$, $x_{rth1} = 0.5$, $x_{rth2} = 0.8$, $k_{r1} = 62 \Omega$, $k_{r2} = 1200 \Omega$, and $k_{r3} = 8440 \Omega$. Within each operating region, the vector field is continuous and the transitions across boundaries preserve

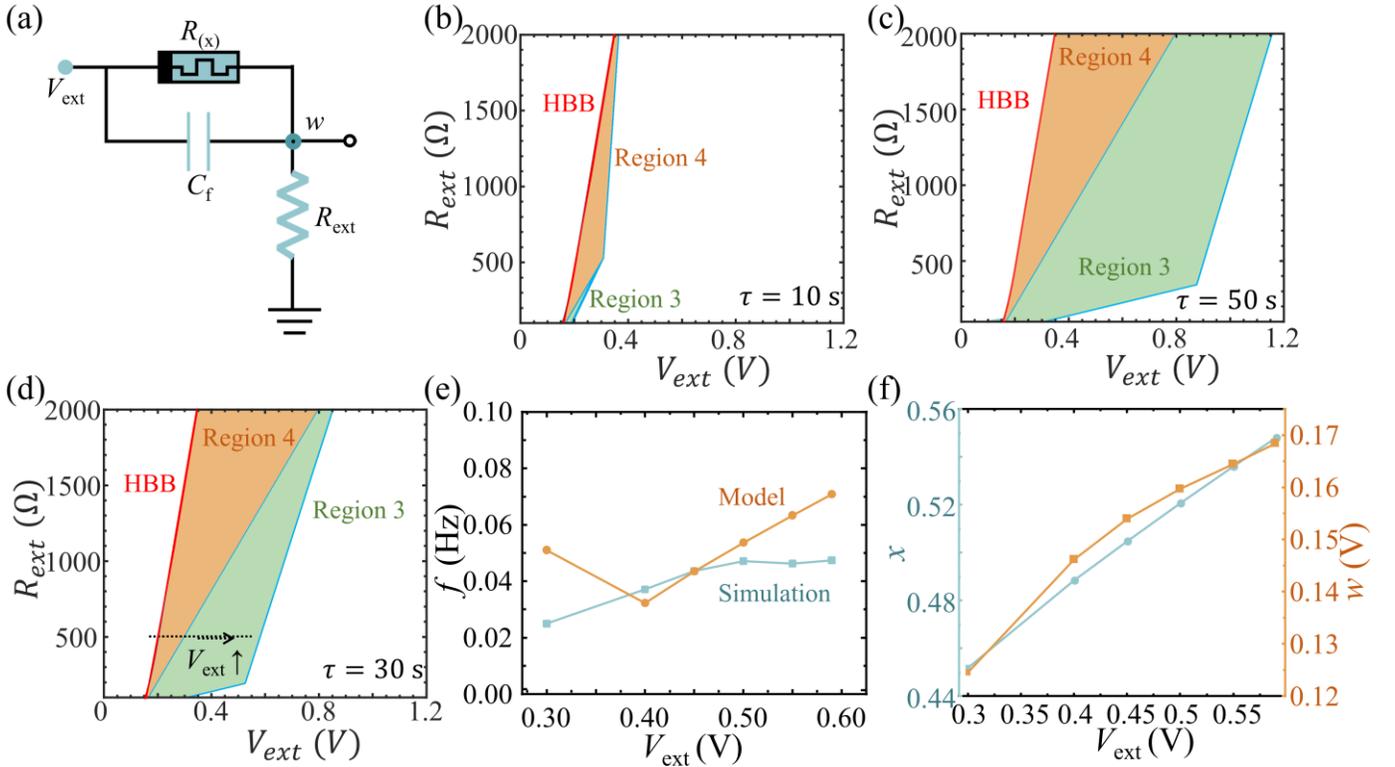

Fig. 4. Dynamic behaviors of the memristor neuron. (a) Neuron circuit schematic realized by involved a memristor. (b–d) Firing domain in ($V_{ext}$, $R_{ext}$) plane with $\tau = 10$ s (b), $\tau = 50$ s (c), and $\tau = 30$ s (d). All other parameters are identical to those used in the main text. The firing domains from different operating regions are shown in different colors. See Note S3 in the SI for the conditions of each operating regions. The red lines mark HBB, and the blue lines mark the boundaries defined by operating regions. The dash line in (d) marks the selected parameters used to assess the effect of $V_{ext}$ on firing properties, that $V_{ext}$ increases from HBB to the interior while $R_{ext}$ is kept constant ($R_{ext} = 500$ Ω). (e) Firing rate changes along $V_{ext}$ from model and simulation. (f) Amplitudes of $x$ and $w$ changes along $V_{ext}$.

trajectory continuity and smoothness.

For the memristor neuron circuit, an important design question is how the device properties, such as volatility speed of the memristor, reflected as $\tau$, influence the FD in the parameter space ($V_{ext}$, $R_{ext}$). By following the developed analysis framework, $\tau$ is assigned three values of 10 s, 30 s, and 50 s, while all other parameters are kept constant to facilitate calculations of the FD in on the ($V_{ext}$, $R_{ext}$) plane. The FD results are displayed in Fig. 4(b)-(d). One boundary of the FD determined by HBB is shown as red lines in Fig. 4(b)-(d). The rest of the boundaries are defined by the corresponding device modes, marked as blue lines. With increasing $\tau$, the FD expansion in the studied range is clearly seen. The FD on the parameter planes of other parameter combinations can be found in Fig. S8 of the SI.

For $\tau = 10$ s (Fig. 4(b)), the memristor state relaxes quickly. Consequently, the state variable $x$ cannot step in the mode of memristor, $R(x)$, with the steepest slop ($x_{th1} < x < x_{th2}$). Fire is then confined to a narrow range of $V_{ext}$ close to HBB. If $\tau$ is small, the state trajectory approaches $w = V_{ext}$, $x = 0$, or $x = 1$ and the neuron can no longer fire.

Increasing $\tau$ to 50 s (Fig. 4(c)) slows down the memristor relaxation, thereby allowing the trajectories to spend more time in the range of $x$ with larger values. The memristor effectively relaxes. Thus, the corresponding FD in ($V_{ext}$, $R_{ext}$) extends over a broader range of drive conditions. Naturally, firing rate decreases with increasing $\tau$ (see Fig. S9 of the SI).

At $\tau = 30$ s, the spike characteristics are investigated for the dependence on $V_{ext}$. As predicted by the model, the firing rate decreases with decreasing $V_{ext}$ from HBB into the interior of the FD, as shown as the blue dot-line in Fig. 4(e). The simulation results are also included as the orange dot-line in the same figure for comparison. Close to HBB, the simulated firing rate agrees well with the model result, confirming that the linearization of the model near the fixed point can capture the firing rate. Away from HBB, however, the simulated firing rate grows faster than the model prediction. The trajectory spends longer in parts of the phase portrait where the vector field is weak and the motion is slow, and, hence, it approaches the boundaries of the allowed states of the system. The behavior of the linearized system deviates from the original system with the state going away from the corresponding fixed point. Thus, the trajectory goes more and more away from the fixed point with decreasing $V_{ext}$, which induces larger deviations.

The spike amplitudes of the state $x$ and the output voltage $w$ with varying $V_{ext}$ are shown in Fig. 4(f). Close to HBB, the two amplitudes scale nearly proportionally with increasing $V_{ext}$, controlled by the local linearized system at the fixed point. When $V_{ext}$ is reduced further, the voltage swing grows slightly faster than the state swing because the load $R_{\text{ext}}$ amplifies the sensitivity of $w$ to $x$ through the state-dependent resistance $R(x)$. In particular, once $x$ traverses the large-slope segment (*i.e.*, $x_{\text{th1}} < x < x_{\text{th2}}$ ), the effective gain $| dw/dx |$ increases, leading to a comparatively large increase in $w$.

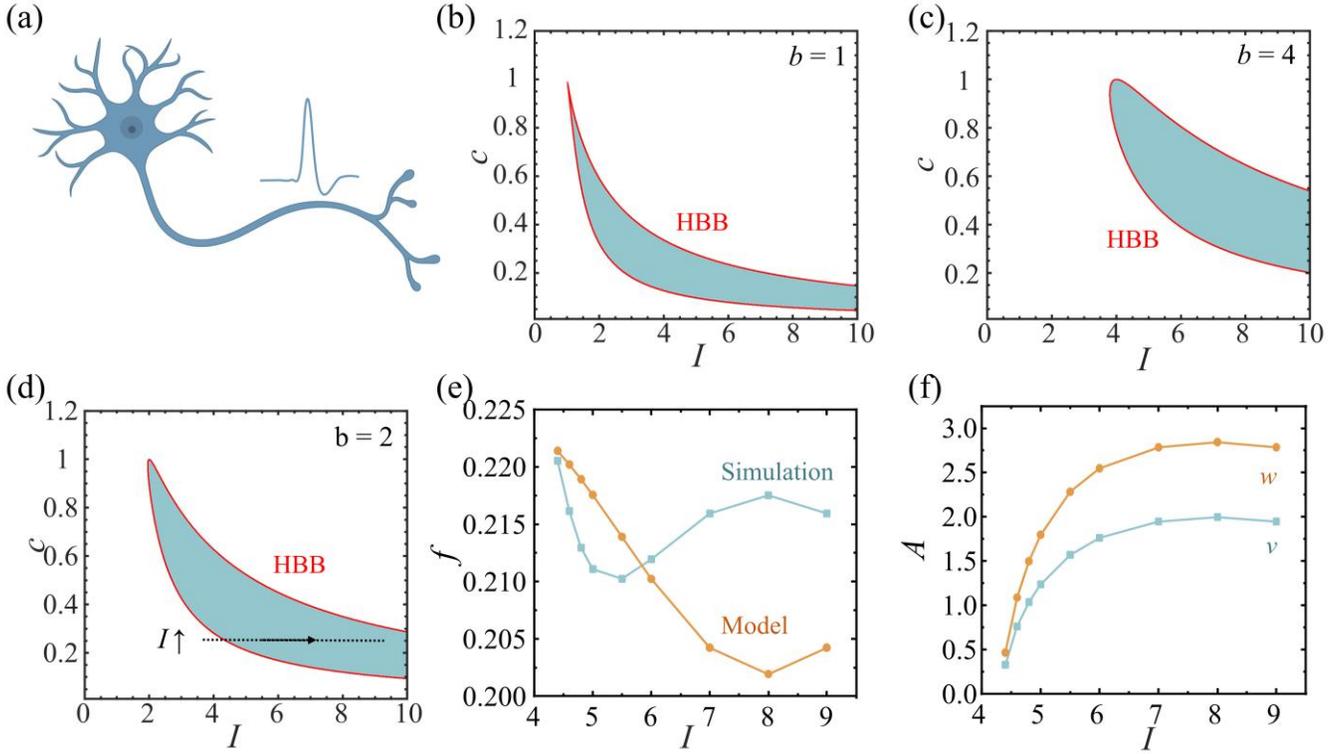

**Fig. 5.** Dynamic behaviors of the FHN model for biological neurons. (a) Schematic illustrating a firing biological neuron. (b–d) Firing domain in $(I, c)$ plane with $b = 1$, $b = 4$, and $b = 2$, respectively. All other parameters are identical to those used in the main text. Red curves mark HBB. The dash line in (d) marks the selected parameters used to assess the effect of $I$ on firing properties, that $I$ increases from HBB to the interior while $c$ is kept constant ($c = 0.25$). (e) Firing rate changes along $I$ from model and simulation. (f) Amplitudes of $v$ and $w$ changes along $I$.

### 3.3 Biological neuron

Below, the same analysis framework is applied with the nondimensional FHN model,

$$\frac{dv}{dt} = v(a - v)(v - 1) - w + I$$
$$\frac{dw}{dt} = bv - cw \tag{16}$$

for biological neurons. Here, $v$ is the membrane potential, $w$ the recovery variable representing sodium and potassium channel activity, $I$ the drive current, and $a$, $b$, and $c$, respectively, offset, recovery gain, and time-scale ratio. In total, the involved parameters of FHN are $\theta = \{a, b, c, I\}$ (see Note 4 of the SI for more details).

In the parameter space, the FD on the $(I, c)$ plane can be identified by varying $I$ and $c$ while keeping the rest parameters invariant, as shown in Fig. 5(b)-(d). Since the system does not have components for different operating modes, it possesses a single operating region. Therefore, the FD in the parameter space is solely determined by Hopf bifurcation conditions, and, thus, its boundary is HBB only, marked as red in Fig. 5(b)-(d). The lower-bound boundary of the FD in the parameter space is defined by the conditions that the fixed point loses its stability, while the upper-bound boundary is defined by the conditions that the fixed point becomes stable again.

The FD shifts toward larger $I$ on the $(I, c)$ plane upon increasing $b$, see the successive changes in Fig. 5(b)-(d). This shift is caused by the $w$-nullcline being tilted upon increasing $b$, leading to a raised $I$ required for the firing onset (see also Fig. S10 of the SI). The shift can reflect an accelerated recovery of the membrane potential back to that of the initial state via stronger feedback after a spike is generated.

To evaluate the dependence of firing rate and spike waveform on $I$, $a = 2$, $b = 2$, and $c = 0.25$ are kept unchanged while $I$ is varied along a line crossing HBB. The firing rate decreases with increasing $I$, as the trajectories would now spend more time in the slowly evolving region in the state space of the system. The firing rate predicted by the model is compared with that of the simulations in Fig. 5(e). Similar to the analysis of the artificial neurons above, the model-predicted firing rate shows closer agreement with the simulation results as $I$ approaches the HBB. The simulated firing rate becomes much higher than the modeled value when $I$ is located deeper in the FD. Farther away from the fixed point, the deviation between the linearized system and the original system grows larger. Furthermore, decreasing $I$ from the HBB enlarges the periodic trajectory, and, thus, the spike amplitude increases, as shown in Fig. 5(f).

*3.4 Discussion*

By demonstrations through the three neuron realizations, *i.e.*, CMOS axon–hillock, memristor, and biological, the proposed framework standardizes the analysis of models and yields a consistent interpretation of spiking dynamics. Although these systems differ in physical origin, their firing is generated from phenomenologically similar elements: fast- and slow-changing nodes with negative feedback (Fig. 1(c)). In the axon–hillock neuron, the slow-changing variable is the input node voltage $v$ whose characteristic relaxation is primarily set by the effective membrane capacitance $(C_{mem} + C_f)$ together with the leak conductance $g_L$. The fast-changing variable is the output-node voltage $w$, which relaxes following the amplifier characteristic $w_\infty$ with time constant $\tau_A$. Negative feedback is implemented by the feedback branch from $w$ back to the input node through the capacitor $C_f$ and the feedback transistor current $I_{fb}$. In the memristor neuron, the voltage of the output node, $w$, evolves rapidly while the internal memristor state, $x$, changes slowly and modulates the conductance. The output voltage is feedback to node $x$ with a quantity of $-qw$ in the first expression of (13). In the FHN model for biological neurons, the membrane potential $v$ is the fast-changing variable while the recovery $w$ is the slow-changing variable. Negative feedback with a quantity of $-w$ connects the input and output nodes to repolarize the neuron, as shown in the first expression of (16). A detailed analysis of the overview structure of the neurons can be found in Note 5 of the SI.

The developed analysis framework here adopts the segmented low-order polynomial function to approximate the often very sophisticated functions for the device behavior, which may well lead to errors. However, these errors can be controlled by increasing the number of segmentations and/or polynomial orders. Such an increase does not necessarily change the trajectory topologically (see Fig. S11 of the SI). It is worth noting that the Lyapunov condition may fail if the order of the polynomial approximation functions is chosen too low.

The spike waveform can also be influenced by the operating region partition in the state space. The spike amplitude corresponds to the maximal excursion of the system state, represented by the trajectory in the state space and permitted by the vector field. Moreover, the larger the intensity of the vector field, the more rapid is the rising and falling edge of the spike waveform. In the CMOS axon–hillock and memristor neurons, the device parameters, $\theta$, define the boundaries of different operating regions in the state space as well as the slopes of the nullclines. Therefore, the waveform can be tuned by adequately choosing these parameters in the corresponding circuits.

## IV. CONCLUSIONS

This work presents a generalized bifurcation analysis framework that is built based on nonlinear dynamical systems analysis. To attain analytical results, the complicated functions of the device/element behaviors in different modes are approximated by segmented low-order polynomial functions. Naturally, the system state space is divided into different operating regions and the analysis is separately implemented in each region. By evaluating the stability and Hopf bifurcation conditions at each fixed point, the FDs and RDs in the parameter space can be defined. In addition, the numerical simulation verifies the firing properties predicted by the model, and further displays the details of the system trajectories in the state space and the corresponding time domain waveforms. To demonstrate the generality, the analysis framework has been applied to three representative application examples: an axon–hillock neuron, a memristor neuron, and the FHN model for biological neurons. It reveals a constant picture of spiking dynamics despite of the vastly different nature of the studied neuron cases. This analysis framework can be extended to higher-dimensional neuron models, by incorporating noise and variation of device characteristics used in the corresponding circuits.

The generalized bifurcation analysis framework of neuron models provides a unified explanation for the onset and termination of firing, evolution of system trajectory, and waveform of spikes for various neuron models. Practically, it guides the parameter selection in device/circuit design and optimization by providing an explicit link between the complicated parameter combinations and neuron behaviors. Thus, it can replace the traditional ad-hoc parameter sweeping method in design. Moreover, it can assist the development of electronic design automation tools dedicated to neuron circuit design.

## ACKNOWLEDGMENT

This work was partially financed by the faculty funds at Uppsala University.

**Zhiwei Li** received the B.Sc. and M.Sc. degrees in precision instrument from Tianjin University, Tianjin, in 2022 and 2024, respectively. He is currently pursuing the Ph.D. degree with Uppsala University, Sweden. His research project concerns novel electronic devices for neuromorphic computing and neuronal dynamic modeling.

**Shi-Li Zhang** received the B.Sc. and M.Sc. degrees in physics from Fudan University, Shanghai, in 1982 and 1985, respectively, and the Ph.D. degree in electronics from the Royal Institute of Technology (KTH), Stockholm, in 1990. He is the Chair Professor of Solid-State Electronics with Uppsala University, Uppsala, Sweden. His current research interests include nanodevices for genome sequencing, flexible energy devices based on 2D semiconductor materials, and energy-efficient interconnects for nano- and power electronics.

**Chenyu Wen** received the B.S. degree in electronic science and technology form Southeast University, Nanjing, China, in 2012, and the Ph.D. degree in electronics from Uppsala University, Sweden, in 2019. He is Assistant Professor of solid-state electronics with Uppsala University, Uppsala, Sweden. His research interests include neuromorphic devices and circuits, electronic device characterization and modeling, nanopore sensors, electrokinetic and ionic devices, gas sensors, and signal processing.

**Supporting Information**

# Bifurcation Analysis Framework of Spiking Neuron Models

**Zhiwei Li[1], Shi-Li Zhang[1], Chenyu Wen[1]\***

[1] *Division of Solid-State Electronics, Department of Electrical Engineering, Uppsala University, SE-751 21 Uppsala, Sweden*

**\*Corresponding author:** <u>chenyu.wen@angstrom.uu.se</u>

**Table of Contents:**

**Supporting Note 1: Derivation of axon–hillock neuron model**

Applying Kirchhoff's current law at the input node and Kirchhoff's voltage law around the feedback loop yields the two-dimensional (2D) axon–hillock model:

$$C_{\text{mem}} \frac{dv}{dt} = I_{\text{in}} - g_{\text{L}} \cdot v + C_{\text{f}} \cdot \frac{d(w-v)}{dt} - I_{\text{fb}}(v, w, \theta_{\text{M}})$$

$$\tau_{\text{A}} \frac{dw}{dt} = w_{\infty}(v, \theta_{\text{A}}) - w$$

(S1)

Rearranging Eq. (S1), $\frac{d(w-v)}{dt} = \frac{dw}{dt} - \frac{dv}{dt}$ gives,

$$(C_{\text{mem}} + C_{\text{f}}) \frac{dv}{dt} = I_{\text{in}} - g_{\text{L}} \cdot v + \frac{C_{\text{f}}}{\tau_{\text{A}}} \cdot (w_{\infty}(v, \theta_{\text{A}}) - w) - I_{\text{fb}}(v, w, \theta_{\text{M}})$$

$$\tau_{\text{A}} \frac{dw}{dt} = w_{\infty}(v, \theta_{\text{A}}) - w$$

(S2)

To non-dimensionalize parameters, we normalized variables by using voltage $V_{\text{dd}}$, time $\tau_{\text{A}}$, and unit charge $q$ ($1.6 \times 10^{-19}$ [C]). $\hat{v} = \frac{v}{V_{\text{dd}}}$, $\hat{w} = \frac{w}{V_{\text{dd}}}$, $\hat{t} = \frac{t}{\tau_{\text{A}}}$, $\hat{C_{\text{f}}} = \frac{C_{\text{f}} V_{\text{dd}}}{q}$, $\hat{I_{\text{in}}} = \frac{I_{\text{in}} \tau_{\text{A}}}{q}$, $\hat{g_{\text{L}}} = \frac{g_{\text{L}} \tau_{\text{A}} V_{\text{dd}}}{q}$, $\hat{V_{\text{th1}}} = \frac{V_{\text{th1}}}{V_{\text{dd}}}$, $\hat{V_{\text{th2}}} = \frac{V_{\text{th2}}}{V_{\text{dd}}}$, $\hat{V_{\text{gth}}} = \frac{V_{\text{gth}}}{V_{\text{dd}}}$, $\hat{k} = \frac{k \tau_{\text{A}} V_{\text{dd}}^2}{q}$.

Dividing the state space into different operating regions by the three modes of the transistor and the amplifier, each with associated constraints, the normalized piecewise model is:

Region 1: $\hat{v} \geq 0$, $\hat{v} \leq \hat{w} - \hat{V_{\text{gth}}}$, $\hat{V_{\text{th2}}} \geq \hat{v} \geq \hat{V_{\text{th1}}}$

$$\frac{d\hat{v}}{d\hat{t}} = \frac{1}{2\hat{C_{\text{f}}}} (\hat{I_{\text{in}}} - \hat{g_{\text{L}}} \hat{v} - \left(\hat{k}(\hat{w} - \hat{V_{\text{gth}}})\hat{v} - \frac{\hat{k}}{2}\hat{v}^2\right) + \hat{C_{\text{f}}}(\frac{\hat{v} - \hat{V_{\text{th1}}}}{(\hat{V_{\text{th2}}} - \hat{V_{\text{th1}}})} - \hat{w}))$$

$$\frac{d\hat{w}}{d\hat{t}} = \frac{\hat{v} - \hat{V_{\text{th1}}}}{(\hat{V_{\text{th2}}} - \hat{V_{\text{th1}}})} - \hat{w}$$

The fixed point $(\hat{v}^*, \hat{w}^*)$ is

$$\hat{v}^* = \frac{\hat{k}\left(\frac{\hat{V_{\text{th1}}}}{\Delta \hat{V}} + \hat{V_{\text{gth}}}\right) - \hat{g_{\text{L}}} \pm \sqrt{\left[\hat{g_{\text{L}}} - \hat{k}\left(\frac{\hat{V_{\text{th1}}}}{\Delta \hat{V}} + \hat{V_{\text{gth}}}\right)\right]^2 + 4\hat{k}(\frac{1}{\Delta \hat{V}} - \frac{1}{2})\hat{I_{\text{in}}}}}{2(\frac{\hat{k}}{\Delta \hat{V}} - \frac{\hat{k}}{2})}$$

$$\hat{w}^* = \frac{\hat{v}^* - \hat{V_{\text{th1}}}}{\Delta \hat{V}}$$

with $\Delta \hat{V} = \hat{V_{\text{th2}}} - \hat{V_{\text{th1}}}$.

Region 2: $\hat{w} \geq \hat{V_{\text{gth}}}$, $\hat{v} > \hat{w} - \hat{V_{\text{gth}}}$, $\hat{V_{\text{th2}}} \geq \hat{v} \geq \hat{V_{\text{th1}}}$

$$\frac{d\hat{v}}{d\hat{t}} = \frac{1}{2\hat{C_{\text{f}}}} (\hat{I_{\text{in}}} - \hat{g_{\text{L}}} \hat{v} - \frac{\hat{k}}{2}(\hat{w} - \hat{V_{\text{gth}}})^2 + \hat{C_{\text{f}}}(\frac{\hat{v} - \hat{V_{\text{th1}}}}{(\hat{V_{\text{th2}}} - \hat{V_{\text{th1}}})} - \hat{w}))$$

$$\frac{d\hat{w}}{d\hat{t}} = \frac{\hat{v} - \hat{V_{\text{th1}}}}{(\hat{V_{\text{th2}}} - \hat{V_{\text{th1}}})} - \hat{w}$$

The fixed point $(\hat{v}^*, \hat{w}^*)$ is

$$\hat{v}^* = \frac{-\left[\frac{\hat{k}}{\Delta \hat{V}}\left(\frac{\widehat{V_{\text{th1}}}}{\Delta \hat{V}} + \widehat{V_{\text{gth}}}\right) - \widehat{g_L}\right] \pm \sqrt{\left[\frac{\hat{k}}{\Delta \hat{V}}\left(\frac{\widehat{V_{\text{th1}}}}{\Delta \hat{V}} + \widehat{V_{\text{gth}}}\right) - \widehat{g_L}\right]^2 + \frac{2\hat{k}}{\Delta \hat{V}^2}\left[\widehat{I_{\text{in}}} - \frac{\hat{k}}{2}\left(\frac{\widehat{V_{\text{th1}}}}{\Delta \hat{V}} + \widehat{V_{\text{gth}}}\right)^2\right]}}{-\frac{\hat{k}}{\Delta \hat{V}^2}}$$

$$\hat{w}^* = \frac{\hat{v}^* - \widehat{V_{\text{th1}}}}{\Delta \hat{V}}$$

Region 3: $\hat{w} \geq 0, \hat{w} < \widehat{V_{\text{gth}}}, \widehat{V_{\text{th2}}} \geq \hat{v} \geq \widehat{V_{\text{th1}}}$

$$\frac{\mathrm{d}\hat{v}}{\mathrm{d}\hat{t}} = \frac{1}{2\hat{C}_f}(\widehat{I_{\text{in}}} - \widehat{g_L}\hat{v} + \hat{C}_f(\frac{\hat{v} - \widehat{V_{\text{th1}}}}{(\widehat{V_{\text{th2}}} - \widehat{V_{\text{th1}}})} - \hat{w}))$$

$$\frac{\mathrm{d}\hat{w}}{\mathrm{d}\hat{t}} = \frac{\hat{v} - \widehat{V_{\text{th1}}}}{(\widehat{V_{\text{th2}}} - \widehat{V_{\text{th1}}})} - \hat{w}$$

The fixed point $(\hat{v}^*, \hat{w}^*)$ is

$$\hat{v}^* = \frac{\widehat{I_{\text{in}}}}{\widehat{g_L}}$$

$$\hat{w}^* = \frac{\hat{v}^* - \widehat{V_{\text{th1}}}}{\Delta \hat{V}}$$

Region 4: $\hat{v} \geq 0, \hat{v} \leq \hat{w} - \widehat{V_{\text{gth}}}, \hat{v} < \widehat{V_{\text{th1}}}$

$$\frac{\mathrm{d}\hat{v}}{\mathrm{d}\hat{t}} = \frac{1}{2\hat{C}_f}(\widehat{I_{\text{in}}} - \widehat{g_L}\hat{v} - \left(\hat{k}(\hat{w} - \widehat{V_{\text{gth}}})\hat{v} - \frac{\hat{k}}{2}\hat{v}^2\right) - \hat{C}_f\hat{w})$$

$$\frac{\mathrm{d}\hat{w}}{\mathrm{d}\hat{t}} = -\hat{w}$$

There is no valid fixed point in this region.

Region 5: $\hat{w} \geq \widehat{V_{\text{gth}}}, \hat{v} > \hat{w} - \widehat{V_{\text{gth}}}, \hat{v} < \widehat{V_{\text{th1}}}$

$$\frac{\mathrm{d}\hat{v}}{\mathrm{d}\hat{t}} = \frac{1}{2\hat{C}_f}(\widehat{I_{\text{in}}} - \widehat{g_L}\hat{v} - \frac{\hat{k}}{2}\left(\hat{w} - \widehat{V_{\text{gth}}}\right)^2 - \hat{C}_f\hat{w})$$

$$\frac{\mathrm{d}\hat{w}}{\mathrm{d}\hat{t}} = -\hat{w}$$

There is no valid fixed point in this region.

Region 6: $\hat{v} \geq 0, \hat{w} < \widehat{V_{\text{gth}}}, \hat{v} < \widehat{V_{\text{th1}}}$

$$\frac{\mathrm{d}\hat{v}}{\mathrm{d}\hat{t}} = \frac{1}{2\hat{C}_f}(\widehat{I_{\text{in}}} - \widehat{g_L}\hat{v} - \hat{C}_f\hat{w})$$

$$\frac{\mathrm{d}\hat{w}}{\mathrm{d}\hat{t}} = -\hat{w}$$

There is no valid fixed point in this region.

Region 7: $\hat{v} \geq 0, \hat{v} \leq \hat{w} - \widehat{V_{\text{gth}}}, \hat{v} > \widehat{V_{\text{th2}}}$

$$\frac{d\hat{v}}{d\hat{t}} = \frac{1}{2\hat{C}_f}\left(\widehat{I_{in}} - \widehat{g_L}\hat{v} - \left(\hat{k}\left(\hat{w} - \widehat{V_{gth}}\right)\hat{v} - \frac{\hat{k}}{2}\hat{v}^2\right) + \hat{C}_f(1 - \hat{w})\right)$$

$$\frac{d\hat{w}}{d\hat{t}} = 1 - \hat{w}$$

The fixed point $(\hat{v}^*, \hat{w}^*)$ is

$$\hat{v}^* = \frac{\left[\widehat{g_L} + \hat{k}\left(1 - \widehat{V_{gth}}\right)\right] \pm \sqrt{\left[\widehat{g_L} + \hat{k}\left(1 - \widehat{V_{gth}}\right)\right]^2 - 2\hat{k}\widehat{I_{in}}}}{\hat{k}}$$

$$\hat{w}^* = 1$$

Region 8: $\hat{w} \geq \widehat{V_{gth}}$ , $\hat{v} > \hat{w} - \widehat{V_{gth}}$, $\hat{v} > \widehat{V_{th2}}$

$$\frac{d\hat{v}}{d\hat{t}} = \frac{1}{2\hat{C}_f}\left(\widehat{I_{in}} - \widehat{g_L}\hat{v} - \frac{\hat{k}}{2}\left(\hat{w} - \widehat{V_{gth}}\right)^2 + \hat{C}_f(1 - \hat{w})\right)$$

$$\frac{d\hat{w}}{d\hat{t}} = 1 - \hat{w}$$

The fixed point $(\hat{v}^*, \hat{w}^*)$ is

$$\hat{v}^* = \frac{\widehat{I_{in}} - \frac{\hat{k}}{2}\left(1 - \widehat{V_{gth}}\right)^2}{\widehat{g_L}}$$

$$\hat{w}^* = 1$$

Region 9: $\hat{v} \geq 0$, $\hat{w} < \widehat{V_{gth}}$, $\hat{v} > \widehat{V_{th2}}$

$$\frac{d\hat{v}}{d\hat{t}} = \frac{1}{2\hat{C}_f}\left(\widehat{I_{in}} - \widehat{g_L}\hat{v} + \hat{C}_f(1 - \hat{w})\right)$$

$$\frac{d\hat{w}}{d\hat{t}} = 1 - \hat{w}$$

The fixed point $(\hat{v}^*, \hat{w}^*)$ is

$$\hat{v}^* = \frac{\widehat{I_{in}}}{\widehat{g_L}}$$

$$\hat{w}^* = 1$$

**Supporting Note 2: Hopf bifurcation conditions**

Consider the 2D neuron model written in vector form as

$$\dot{x} = F(x, \theta)$$

with $x = \begin{bmatrix} v \\ w \end{bmatrix}, F(x, \theta) = \begin{bmatrix} f(x, \theta) \\ g(x, \theta) \end{bmatrix}, \dot{x} = \begin{bmatrix} dv/dt \\ dw/dt \end{bmatrix},$

where, $\theta$ denotes the set of system parameters. Because the model is piecewise-defined by operating regions in the $(v, w)$ state space, the following Hopf bifurcation conditions are evaluated within each region on each fixed point that lies inside the corresponding operating region.

1. For each operating region, compute fixed points $x^* = (v^*, w^*)$ by solving the nullcline intersection conditions $\mathcal{N}_v = 0$ and $\mathcal{N}_w = 0$. At least one valid fixed point must exist in the region $\Omega_k, k = 1 \dots 9$.

2. At each fixed point in corresponding region, form the Jacobian matrix

$$J(x^*, \theta) = \frac{dF}{dx}\bigg|_{x=x^*}$$

A Hopf bifurcation requires that $J(x^*, \theta)$ has a complex-conjugate eigenvalue pair

$$\lambda_{1,2} = a(\theta) \pm i\omega(\theta), \omega \neq 0$$

For a 2D system, the trace and determinant satisfy

$$Tr = tr(J) = 2a, \Delta = \det(J) = a^2 + \omega^2$$

3. Let $\theta_i$ be the parameter chosen for continuation. A candidate Hopf point $\theta_i = \theta_{i,0}$ is obtained by solving

$$Tr(\theta_i) = 2a(\theta_i) = 0$$

subject to

$$\Delta(\theta_{i,0}) > 0$$

which ensures purely imaginary eigenvalues $\lambda_{1,2} = \pm i\omega_0$ at the Hopf bifurcation $\theta_i = \theta_{i,0}$.

4. To guarantee a Hopf bifurcation rather than a degenerate case, the eigenvalues must cross the imaginary axis with nonzero "speed" as $\theta_i$ varies,

$$\frac{dRe(\lambda)}{d\theta_i}\bigg|_{\theta_i = \theta_{i,0}} = \frac{da}{d\theta_i}\bigg|_{\theta_i = \theta_{i,0}} \neq 0$$

5. To determine whether the Hopf bifurcation generates stable small-amplitude oscillations, we compute the first Lyapunov coefficient $Lv$ at $\theta_i = \theta_{i,0}$, and require $Lv < 0$.

$$Lv = \frac{1}{16}(f_{vvv} + f_{vww} + g_{vvw} + g_{www})$$

$$+ \frac{1}{16\omega}(f_{vw}(f_{vv} + f_{ww}) - g_{vw}(g_{vv} + g_{ww}) - f_{vv}g_{vv} + f_{ww}g_{ww})$$

6. In summary, these conditions give a group of inequality of the parameters that makes a neuron fire

$$x^* \in \Omega_k$$

$$Tr\big(J(x^*, \theta_{i,0})\big) = 0$$

$$\Delta\big(J(x^*, \theta_{i,0})\big) > 0$$

$$\left.\frac{dRe(\lambda)}{d\theta_i}\right|_{\theta_i = \theta_{i,0}} \neq 0$$

$$Lv(x^*, \theta_{i,0}) < 0.$$

For illustration, we take region 1 in the CMOS axon–hillock neuron model as an example. If $\widehat{I_{\text{in}}}$ is chosen as the interested parameter. The valid fixed point $\hat{x}^* = (\hat{v}^*, \hat{w}^*)$ in region 1 is

$$\hat{v}^* = \frac{\hat{k}\left(\frac{\widehat{V_{\text{th1}}}}{\Delta \hat{V}} + \widehat{V_{\text{gth}}}\right) - \widehat{g_{\text{L}}} \pm \sqrt{\left[\widehat{g_{\text{L}}} - \hat{k}\left(\frac{\widehat{V_{\text{th1}}}}{\Delta \hat{V}} + \widehat{V_{\text{gth}}}\right)\right]^2 + 4\hat{k}(\frac{1}{\Delta \hat{V}} - \frac{1}{2})\widehat{I_{\text{in}}}}}{2(\frac{\hat{k}}{\Delta \hat{V}} - \frac{\hat{k}}{2})}$$

$$\hat{w}^* = \frac{\hat{v}^* - \widehat{V_{\text{th1}}}}{\Delta \hat{V}}$$

So, the existence of real-valued fixed point requires:

$$\left[\widehat{g_{\text{L}}} - \hat{k}\left(\frac{\widehat{V_{\text{th1}}}}{\Delta \hat{V}} + \widehat{V_{\text{gth}}}\right)\right]^2 + 4\hat{k}\left(\frac{1}{\Delta \hat{V}} - \frac{1}{2}\right)\widehat{I_{\text{in}}} > 0$$

In Region 1, the valid fixed points satisfy

$$\hat{v}^* \geq 0$$

$$\hat{v}^* \leq \hat{w}^* - \widehat{V_{\text{gth}}}$$

$$\widehat{V_{\text{th2}}} \geq \hat{v}^* \geq \widehat{V_{\text{th1}}}$$

The inequalities from the Hopf bifurcation conditions that define the firing region are:

$$Tr\big(J(\hat{x}^*, \widehat{I_{\text{in}}})\big) = \frac{-\widehat{g_{\text{L}}} - \hat{k}\big(\hat{w}^* - \widehat{V_{\text{gth}}}\big) + \hat{k}\hat{v}^* + \frac{\hat{c}_{\text{f}}}{\Delta \hat{V}}}{2\hat{c}_{\text{f}}} - 1 > 0$$

$$\Delta\big(J(\hat{x}^*, \widehat{I_{\text{in}}})\big) = -\frac{-\widehat{g_{\text{L}}} - \hat{k}\big(\hat{w}^* - \widehat{V_{\text{gth}}}\big) + \hat{k}\hat{v}^* + \frac{\hat{c}_{\text{f}}}{\Delta \hat{V}}}{2\hat{c}_{\text{f}}} - \frac{-\hat{k}\hat{v}^* - \hat{c}_{\text{f}}}{2\Delta \hat{V}\hat{c}_{\text{f}}} > 0$$

$$\left.\frac{dRe(\lambda)}{d\theta_i}\right|_{\theta_i = \widehat{I_{\text{in}}}} = \frac{d\frac{-\widehat{g_{\text{L}}} - \hat{k}\big(\hat{w}^* - \widehat{V_{\text{gth}}}\big) + \hat{k}\hat{v}^* + \frac{\hat{c}_{\text{f}}}{\Delta \hat{V}}}{2\hat{c}_{\text{f}}}}{d\widehat{I_{\text{in}}}} \neq 0$$

$$Lv\big(\hat{x}^*, \widehat{I_{\text{in}}}\big) = \frac{1}{16\omega^*}\frac{-1}{4\hat{c}_{\text{f}}^2} < 0$$

**Supporting Note 3: Derivation of memristor neuron model**

The ODEs of the memristor neurons are written:

$$\frac{\mathrm{d}x}{\mathrm{d}t} = -u(x) - qw$$

$$\tau\frac{\mathrm{d}w}{\mathrm{d}t} = V_{\text{ext}} - \left(1 + \frac{R_{\text{ext}}}{R(x)}\right)w$$

Under the physics constraint $x_{\text{th1}} < x_{\text{rth1}} < x_{\text{rth2}} < x_{\text{th2}}$, the state space is partitioned into five operating regions in $(x, w)$. In each region, $u(x)$ and $R(x)$ in the ODEs are given by the corresponding expressions.

Region 1: $x < x_{\text{th1}}$

$$\frac{\mathrm{d}x}{\mathrm{d}t} = -(xk_1 + x_{\text{u0}}) - qw$$

$$\frac{\mathrm{d}w}{\mathrm{d}t} = \frac{1}{\tau}V_{\text{ext}} - \frac{1}{\tau}\left(1 + \frac{R_{\text{ext}}}{xk_{\text{r1}} + 1}\right)w$$

The fixed point $(x^*, w^*)$ is

$$x^* = \frac{-A_1 \pm \sqrt{A_1{}^2 - B_1}}{2\dfrac{k_{\text{r1}}k_1}{q}}$$

$$w^* = \frac{V_{\text{ext}}}{\left(1 + \dfrac{R_{\text{ext}}}{x^*k_{\text{r1}} + 1}\right)}$$

With

$$A_1 = V_{\text{ext}}k_{\text{r1}} + (1 + R_{\text{ext}})\frac{k_1}{q} + \frac{k_{\text{r1}}x_{\text{u0}}}{q}$$

$$B_1 = 4\frac{k_{\text{r1}}k_1}{q}\left(V_{\text{ext}} + \frac{(1 + R_{\text{ext}})x_{\text{u0}}}{q}\right)$$

Region 2: $x_{th1} < x < x_{rth1}$

$$\frac{\mathrm{d}x}{\mathrm{d}t} = -[(x - x_{\text{th1}})k_2 + x_{\text{th1}}k_1 + x_{\text{u0}}] - qw$$

$$\frac{\mathrm{d}w}{\mathrm{d}t} = \frac{1}{\tau}V_{\text{ext}} - \frac{1}{\tau}\left(1 + \frac{R_{\text{ext}}}{xk_{\text{r1}} + 1}\right)w$$

The fixed point $(x^*, w^*)$ is

$$x^* = \frac{-A_2 \pm \sqrt{A_2{}^2 - B_2}}{2\frac{k_{r1}k_2}{q}}$$

$$w^* = \frac{V_{\text{ext}}}{\left(1 + \frac{R_{\text{ext}}}{x^* k_{r1} + 1}\right)}$$

With

$$m_1 = x_{\text{th1}}k_1 + x_{\text{u0}} - x_{\text{th1}}k_2$$

$$A_2 = V_{\text{ext}}k_{r1} + (1 + R_{\text{ext}})\frac{k_2}{q} + \frac{k_{r1}m_1}{q}$$

$$B_2 = 4\frac{k_{r1}k_2}{q}\left(V_{\text{ext}} + \frac{(1 + R_{\text{ext}})m_1}{q}\right)$$

Region 3: $x_{rth1} < x < x_{rth2}$

$$\frac{\mathrm{d}x}{\mathrm{d}t} = -[(x - x_{\text{th1}})k_2 + x_{\text{th1}}k_1 + x_{\text{u0}}] - qw$$

$$\frac{\mathrm{d}w}{\mathrm{d}t} = \frac{1}{\tau}V_{\text{ext}} - \frac{1}{\tau}\left(1 + \frac{R_{\text{ext}}}{(x - x_{\text{rth1}})k_{r2} + x_{\text{rth1}}k_{r1} + 1}\right)w$$

Fixed point $(x^*, w^*)$

$$x^* = \frac{-A_3 \pm \sqrt{A_3{}^2 - B_3}}{2\frac{k_{r2}k_2}{q}}$$

$$w^* = \frac{V_{\text{ext}}}{\left(1 + \frac{R_{\text{ext}}}{x^* k_{r2} + m_2}\right)}$$

With

$$m_2 = -x_{\text{rth1}}k_{r2} + x_{\text{rth1}}k_{r1} + 1$$

$$A_3 = V_{\text{ext}}k_{r2} + (m_2 + R_{\text{ext}})\frac{k_2}{q} + \frac{k_{r2}m_1}{q}$$

$$B_3 = 4\frac{k_{r2}k_2}{q}\left(V_{\text{ext}}m_2 + \frac{(m_2 + R_{\text{ext}})m_1}{q}\right)$$

Region 4: $x_{rth2} < x < x_{th2}$

$$\frac{\mathrm{d}x}{\mathrm{d}t} = -[(x - x_{\text{th1}})k_2 + x_{\text{th1}}k_1 + x_{\text{u0}}] - qw$$

$$\frac{\mathrm{d}w}{\mathrm{d}t} = \frac{1}{\tau} V_{\text{ext}} - \frac{1}{\tau}\left(1 + \frac{R_{\text{ext}}}{(x - x_{\text{rth2}})k_{\text{r3}} + (x_{\text{rth2}} - x_{\text{rth1}})k_{\text{r2}} + x_{\text{rth1}}k_{\text{r1}} + 1}\right)w$$

The fixed point $(x^*, w^*)$ is

$$x^* = \frac{-A_4 \pm \sqrt{A_4{}^2 - B_4}}{2\frac{k_{\text{r3}}k_2}{q}}$$

$$w^* = \frac{V_{\text{ext}}}{\left(1 + \frac{R_{\text{ext}}}{x^* k_{\text{r3}} + m_3}\right)}$$

With

$$m_3 = -x_{\text{rth2}}k_{\text{r3}} + (x_{\text{rth2}} - x_{\text{rth1}})k_{\text{r2}} + x_{\text{rth1}}k_{\text{r1}} + 1$$

$$A_4 = V_{\text{ext}}k_{\text{r3}} + (m_3 + R_{\text{ext}})\frac{k_2}{q} + \frac{k_{\text{r3}}m_1}{q}$$

$$B_4 = 4\frac{k_{\text{r3}}k_2}{q}\left(V_{\text{ext}}m_3 + \frac{(m_3 + R_{\text{ext}})m_1}{q}\right)$$

Region 5: $x_{\text{th2}} < x$

$$\frac{\mathrm{d}x}{\mathrm{d}t} = -[(x - x_{\text{th2}})k_3 + (x_{\text{th2}} - x_{\text{th1}})k_2 + x_{\text{th1}}k_1 + x_{\text{u0}}] - qw$$

$$\frac{\mathrm{d}w}{\mathrm{d}t} = \frac{1}{\tau} V_{\text{ext}} - \frac{1}{\tau}\left(1 + \frac{R_{\text{ext}}}{(x - x_{\text{rth2}})k_{\text{r3}} + (x_{\text{rth2}} - x_{\text{rth1}})k_{\text{r2}} + x_{\text{rth1}}k_{\text{r1}} + 1}\right)w$$

The fixed point $(x^*, w^*)$ is

$$x^* = \frac{-A_5 \pm \sqrt{A_5{}^2 - B_5}}{2\frac{k_{\text{r3}}k_3}{q}}$$

$$w^* = \frac{V_{\text{ext}}}{\left(1 + \frac{R_{\text{ext}}}{x^* k_{\text{r3}} + m_3}\right)}$$

With

$$m_4 = -x_{\text{th2}}k_3 + (x_{\text{th2}} - x_{\text{th1}})k_2 + x_{\text{th1}}k_1 + x_{\text{u0}}$$

$$A_5 = V_{\text{ext}}k_{\text{r3}} + (m_3 + R_{\text{ext}})\frac{k_3}{q} + \frac{k_{\text{r3}}m_4}{q}$$

$$B_5 = 4\frac{k_{\text{r3}}k_3}{q}\left(V_{\text{ext}}m_3 + \frac{(m_3 + R_{\text{ext}})m_4}{q}\right)$$

**Supporting Note 4: Details of the FHN model**

The FHN model is

$$\frac{dv}{dt} = v(a-v)(v-1) - w + I$$

$$\frac{dw}{dt} = bv - cw \tag{15}$$

where, $v$ is the membrane potential, $w$ is the recovery variable representing sodium and potassium channel activity, $I$ is the drive current, and $a$, $b$, and $c$ are parameters of offset, recovery gain, and time-scale ratio. In total, the involved parameters of FHN are $\theta = \{a, b, c, I\}$. The fixed point $(v^*, w^*)$ is

$$v_k^* = \frac{a+1}{3} + e^{\frac{2}{3}i\pi k}\sqrt[3]{-\frac{p}{2} + \sqrt{\left(\frac{p}{2}\right)^2 + \left(\frac{h}{2}\right)^3}} + e^{-\frac{2}{3}i\pi k}\sqrt[3]{-\frac{p}{2} - \sqrt{\left(\frac{p}{2}\right)^2 + \left(\frac{h}{2}\right)^3}}, k = 0,1,2$$

$$w_k^* = \frac{b}{c}v_k^*, \qquad k = 0,1,2$$

With

$$p = -\frac{2(a+1)^3}{27} + \frac{(a+1)}{3}\left(a + \frac{b}{c}\right) - I$$

$$h = \left(a + \frac{b}{c}\right) - -\frac{(a+1)^3}{3}$$

**Supporting Note 5: A unified system level view of neuron models**

To present the three neuron realizations in a unified and compact manner, we decompose each model into four functional units, (i) time-constant relaxation terms, (ii) an input, (iii) a nonlinear forward path, and (iv) explicit negative feedback branches. In all cases, spiking arises from a closed-loop interaction between fast and slow state variables under this structure.

- Axon–hillock neuron

$$\frac{dv}{dt} = - \underbrace{\frac{g_\mathrm{L}}{C_\mathrm{mem} + C_\mathrm{f}}}_{1/\text{time constant}} v + \underbrace{\frac{I_\mathrm{in}}{C_\mathrm{mem} + C_\mathrm{f}}}_{\text{input}} + \underbrace{\frac{C_\mathrm{f}}{\tau_\mathrm{A}(C_\mathrm{mem} + C_\mathrm{f})}(w_\infty(v,\theta_\mathrm{A}) - w)}_{\text{capacitive feedback}} - \underbrace{\frac{I_\mathrm{fb}(v,w,\theta_\mathrm{T})}{(C_\mathrm{mem} + C_\mathrm{f})}}_{\text{transistor feedback}}$$

$$\frac{dw}{dt} = - \underbrace{\frac{1}{\tau_\mathrm{A}}}_{1/\text{time constant}} w + \underbrace{\frac{w_\infty(v,\theta_\mathrm{A})}{\tau_\mathrm{A}}}_{\text{forward path}}$$

Here $v$ is the slow-changing input-node voltage, with relaxation primarily determined by $C_\mathrm{mem} + C_\mathrm{f}$ and $g_\mathrm{L}$, whereas $w$ is the fast-changing output-node voltage relaxing toward the amplifier characteristic $w_\infty$ with time constant $\tau_\mathrm{A}$. Negative feedback from $w$ to $v$ is realized by the capacitive term proportional to $w_\infty(v,\theta_\mathrm{A}) - w$ and the transistor feedback current $I_\mathrm{fb}(v,w,\theta_\mathrm{T})$.

- Memristor neuron

$$\frac{dx}{dt} = - \underbrace{u(x)}_{1/\text{time constant}} - \underbrace{qw}_{\text{feedback}}$$

$$\frac{dw}{dt} = - \underbrace{\frac{1}{\tau}}_{1/\text{time constant}} w + \underbrace{\frac{V_\mathrm{ext}}{\tau}}_{\text{input}} - \underbrace{\frac{R_\mathrm{ext}}{\tau R(x)}w}_{\text{forward path}}$$

In this realization, $w$ is the fast-changing voltage state, while the internal memristor state $x$ evolves slowly and modulates the conductance through $R(x)$. The term $-qw$ closes the negative-feedback loop by coupling the output activity to the slow state, and the forward path contribution in $\dot w$ depends explicitly on $R(x)$.

- FHN model

$$\frac{dv}{dt} = - \underbrace{a}_{1/\text{time constant}} v + (a+1)v^2 - v^3 - \underbrace{w}_{feedback} + \underbrace{I}_{input}$$

$$\frac{dw}{dt} = - \underbrace{c}_{1/\text{time constant}} w + \underbrace{bv}_{forward\ path}$$

Here $v$ is the fast-changing membrane potential and $w$ is the slow-changing recovery variable. Negative feedback is implemented by the $-w$ term in $\dot v$, while the nonlinear forward path in $\dot v$ provides excitability and the coupling $bv$ drives recovery in $\dot w$.

# Supporting Figures

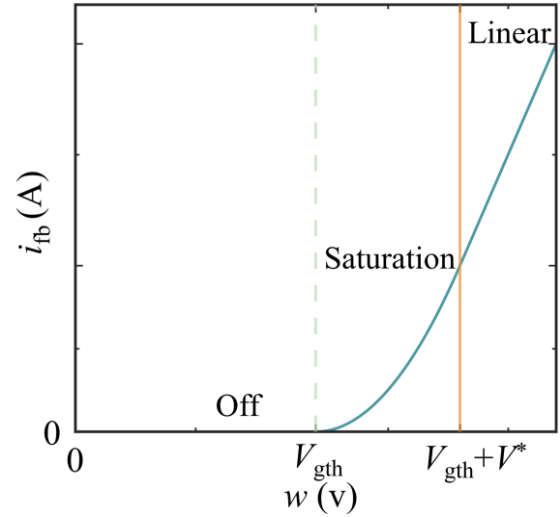

Fig. S1. Plot of $I_{\text{fb}}(v, w, \theta_{\text{T}})$ of the transistor in the axon–hillock model, showing different operating modes, Off, Saturation, and Linear. The parameters used for this curve: $k = 10^{-5}$ A/V$^2$, $V_{\text{gth}} = 1.5$ V, $V_{\text{dd}} = 3.0$ V, $v = 0.9$ V.

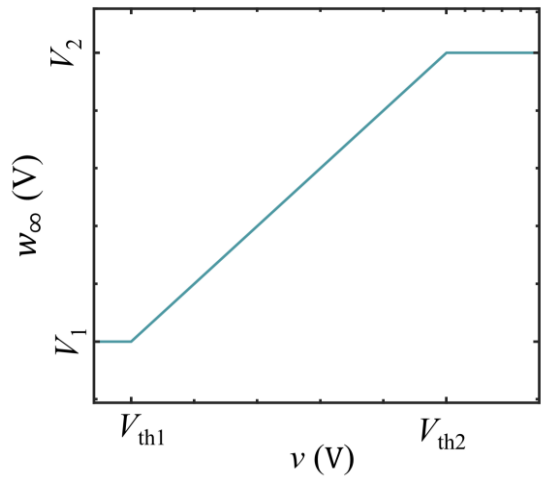

Fig. S2. Plot of the segmented transfer function of the amplifier, $w_\infty(v, \theta_A)$. The parameters used for this curve: $V_{\text{th1}} = 0.1$ V, $V_{\text{th2}} = 0.95$ V, $V_1 = 0.1$ V, $V_2 = 0.95$ V, $V_{\text{dd}} = 1.0$ V.

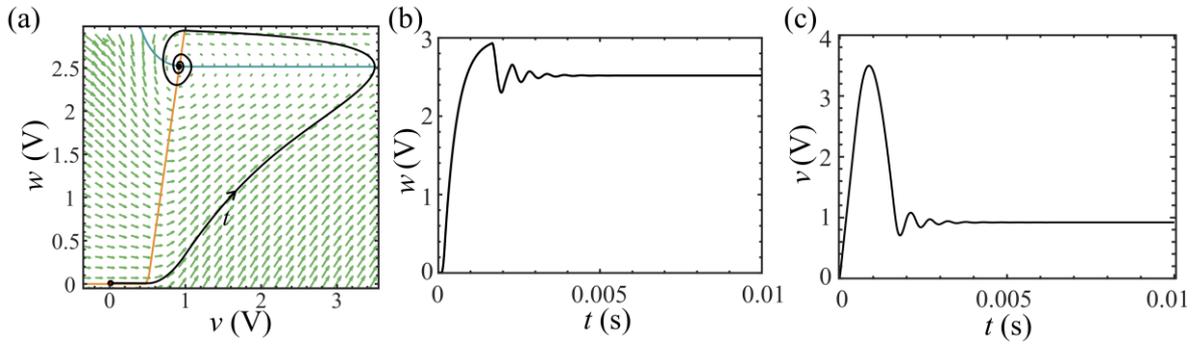

Fig. S3. Phase portrait of the axon–hillock model with $C_\mathrm{f} = 0$ nF, illustrating the rest state. (a) Phase portrait in the $(v, w)$ plane, showing trajectories (black curve) converging to the stable fixed point and the corresponding vector field (arrows) and nullclines (blue-grey and orange curves). (b) Time-domain waveform $v(t)$. (c) Time-domain waveform $w(t)$. All parameters are identical to those used in the main text unless stated otherwise.

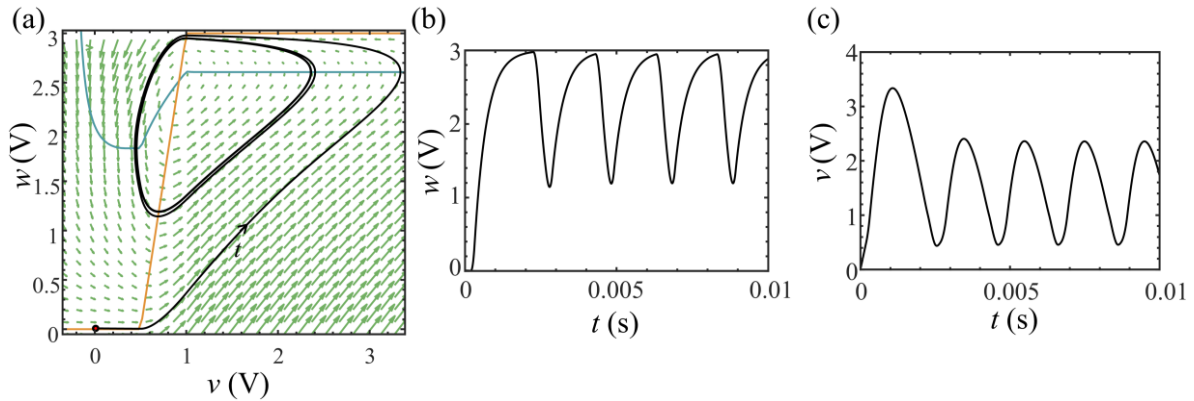

Fig. S4. Phase portrait of the axon–hillock model with $C_\mathrm{f} = 1$ nF in the firing state. (a) Phase portrait in the $(v, w)$ plane, showing the stable limit cycle together with the underlying vector field and nullclines. Black curve is the system trajectory. The blue-grey and orange curves show the nullclines, and the arrows display the vector field. (b) Time-domain waveform $v(t)$ over representative cycles. (c) Time-domain waveform $w(t)$, showing the corresponding output node evolution. All parameters are identical to those used in the main text unless stated otherwise.

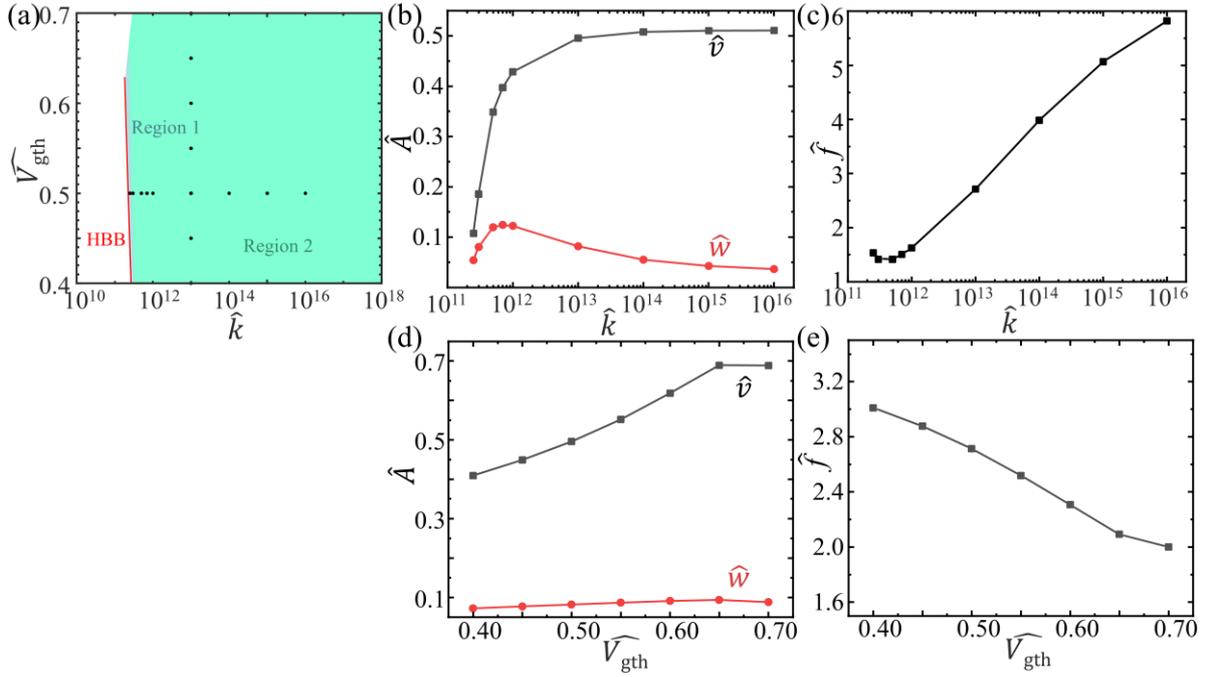

Fig. S5. Parameter dependence of firing of the axon–hillock model. (a) Firing domain in the $\left(\hat{k}, \widehat{V_{\text{gth}}}\right)$ plane, indicating parameter combinations that produce sustained spiking. The firing domains from different operating regions are shown in different colors. See Note 1 in the SI for the conditions of each operating regions. The red line marks HBB. The black dots mark the selected parameters used to assess the effect of $\hat{k}$ and $\widehat{V_{\text{gth}}}$ on firing properties, that $\hat{k}$ increases from HBB to the interior while $\widehat{V_{\text{gth}}} = 0.5$ (horizontal dot sequence) and $\widehat{V_{\text{gth}}}$ increases while $\hat{k}$ is kept constant of $10^{13}$ (vertical dot sequence). (b, c) Spike amplitude and firing rate as functions of $\hat{k}$, evaluated along the horizontally distributed sample points marked in (a). (d, e) Spike amplitude and firing rate as functions of $\widehat{V_{\text{gth}}}$, evaluated along the vertically distributed sample points marked in (a). All other parameters are identical to those used in the main text unless stated otherwise.

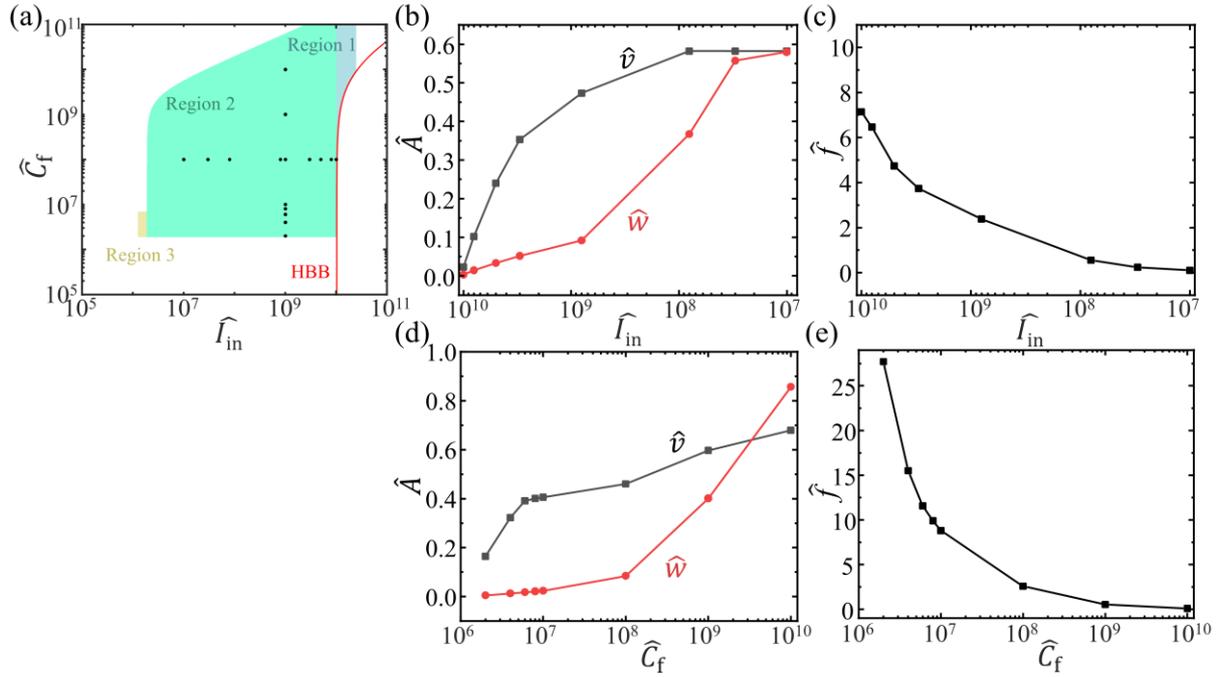

Fig. S6. Parameter dependence of firing properties of the axon–hillock model. (a) Firing domain in the $(\widehat{I_{\mathrm{in}}}, \widehat{C_{\mathrm{f}}})$ plane, indicating parameter combinations that yield sustained spiking. The firing domains from different operating regions are shown in different colors. See Note 1 in the SI for the conditions of each operating regions. The red line marks HBB. The black dots mark the selected parameters used to assess the effect of $\widehat{I_{\mathrm{in}}}$ and $\widehat{C_{\mathrm{f}}}$ on firing properties, that $\widehat{I_{\mathrm{in}}}$ decreases from HBB to the interior while $\widehat{C_{\mathrm{f}}} = 10^8$ (horizontal dot sequence) and $\widehat{C_{\mathrm{f}}}$ increases while $\widehat{I_{\mathrm{in}}}$ is kept constant of $10^9$ (vertical dot sequence). (b, c) Spike amplitude and firing rate as functions of $\widehat{I_{\mathrm{in}}}$, evaluated along the horizontally distributed sample points marked in (a). (d, e) Spike amplitude and firing rate as functions of $\widehat{C_{\mathrm{f}}}$, evaluated along the vertically distributed sample points marked in (a). All other parameters are identical to those used in the main text unless stated otherwise.

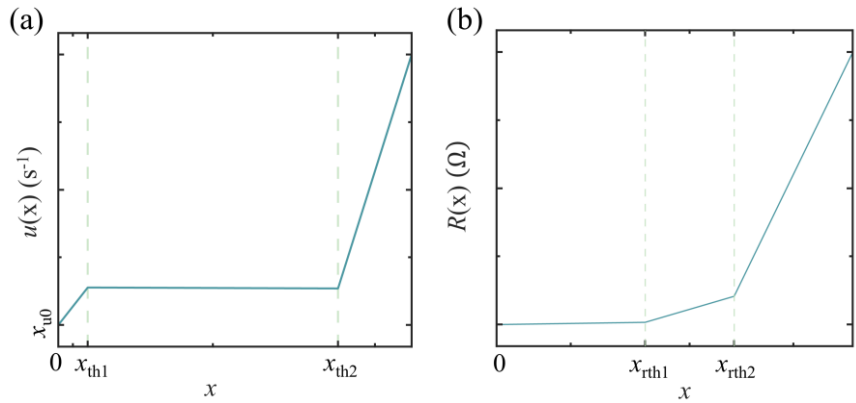

Fig. S7. Plot of the segmented function of $u(x)$ and $R(x)$ in the memristor neuron model. The parameters used for this curve: $x_{\mathrm{th1}} = 0.10$, $x_{\mathrm{th2}} = 0.95$, $k_1 = 80$ V, $k_2 = -0.2353$ V, $k_3 = 200$ V, $x_{\mathrm{u0}} = -8$, $x_{\mathrm{rth1}} = 0.50$, $x_{\mathrm{rth2}} = 0.80$, $k_1 = 62$ V, $k_2 = 1200$ V, $k_3 = 8440$ V.

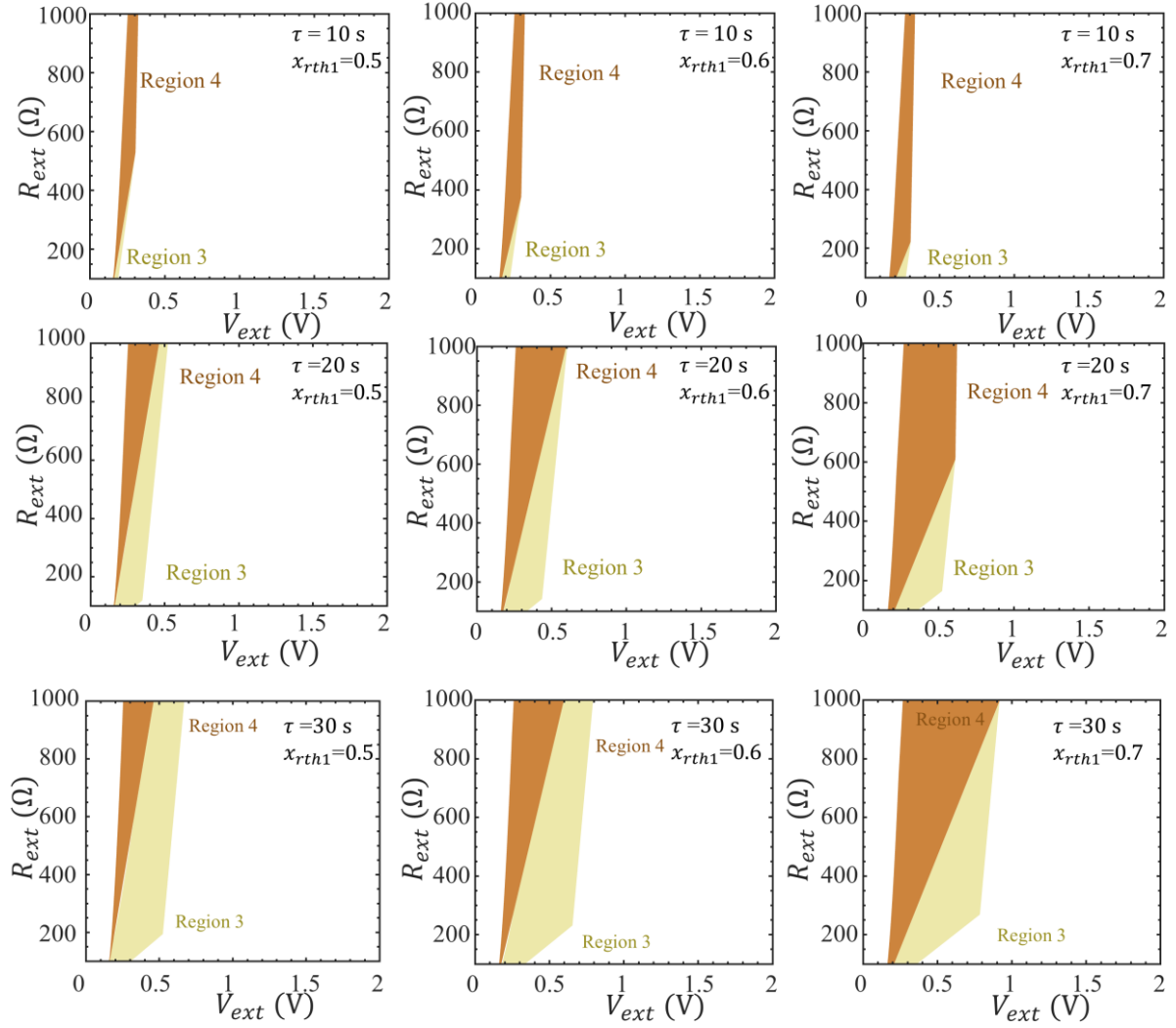

Fig. S8. Firing domain of the memristor neuron model in $(I, c)$ parameter plane with varied $\tau$ and $x_{\text{rth1}}$. The firing domains from different operating regions are shown in different colors. See Note 3 in the SI for the conditions of each operating regions. The other parameters used for this group of figures: $x_{\text{th1}} = 0.1$, $x_{\text{th2}} = 0.95$, $x_{\text{u0}} = -8 \, \text{s}^{-1}$, $k_1 = 80 \, \text{s}^{-1}$, $k_2 = -0.2353 \, \text{s}^{-1}$, $k_3 = 200 \, \text{s}^{-1}$, $q = 1.265 \, \text{V}^{-1}\text{s}^{-1}$, $x_{\text{rth2}} = 0.8$, $\text{k}_{\text{r1}} = 62 \, \Omega$, $\text{k}_{\text{r2}} = 1200 \, \Omega$, and $k_{\text{r3}} = 8440 \, \Omega$.

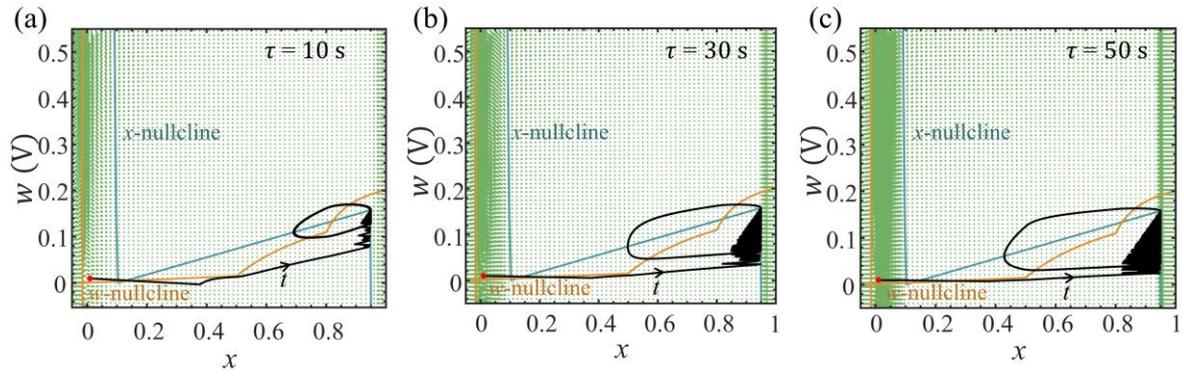

Fig. S9. System trajectories in phase portrait showing a firing state in memristor model with increasing $\tau$ from 10 s (a) and 30 s (b) to 50 s (c). It can be seen that the trajectory reaches a boundary at the right-hand side in the state space ($x = 0.95$) defined by the maximum range of $x$ that can be reached by the memristor device. All other parameters are identical to those used in the main text unless stated otherwise.

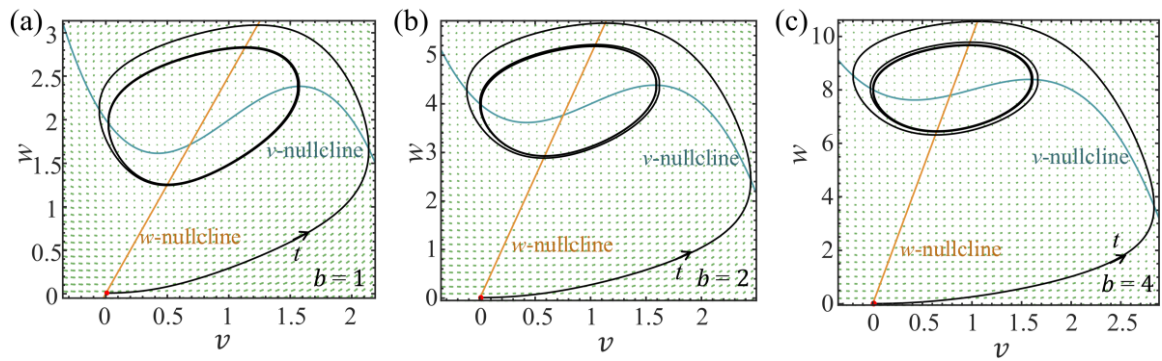

Fig. S10. System trajectories in phase portrait showing a firing state in the FHN model with increasing $b$ from 1 (a) and 2 (b) to 4 (c). The other parameters used for this group of figures: $a = 2$, $I = 2$, and $c = 0.25$.

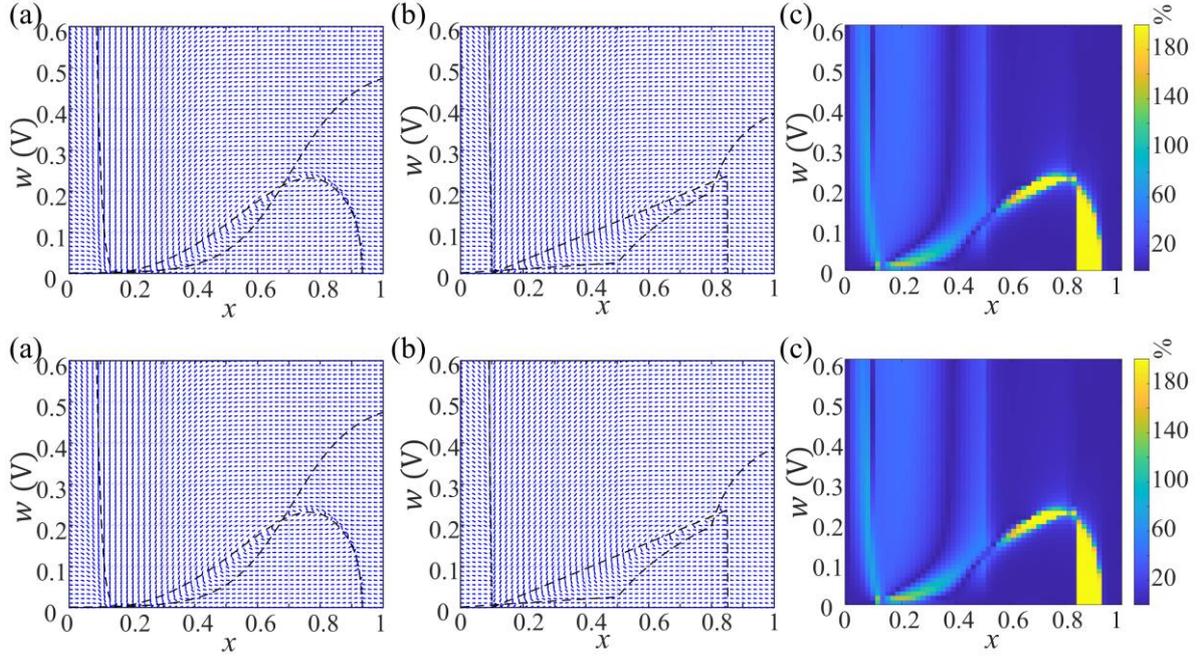

Fig. S11 In the memristor model phase portrait, comparison between the original and piecewise-approximated models. (a) Phase portrait of the original model, showing the vector field $f_o(v,w)$ and the corresponding nullclines. (b) Phase portrait of the piecewise model under identical parameter settings and plotting range, showing the vector field $f_{pw}(v,w)$ and the corresponding nullclines. (c) Heatmap of the pointwise modulus of the phasor discrepancy, $\|\Delta f(v,w)\| = \frac{\|f_{pw}(v,w) - f_o(v,w)\|}{\|f_o(v,w)\|} \times 100\%$, computed from the relative difference between the normalized vector fields from the original and piecewise systems. All other parameters are identical to those used in the main text unless stated otherwise.